%% file: main.tex
\documentclass[fleqn,usenatbib,useAMS]{mnras}

\usepackage{subfiles}
\usepackage{amsmath}	% Advanced maths commands
\usepackage{amssymb}	% Extra maths symbols
\usepackage{multicol}        % Multi-column entries in tables
\usepackage{bm}		% Bold maths symbols, including upright Greek
\usepackage{pdflscape}	% Landscape pages
\usepackage{microtype}      % microtypography
\usepackage{float}
\usepackage{graphicx}
\usepackage{caption}
\usepackage{natbib}
\usepackage{lipsum}
\usepackage{calc}  
\usepackage{enumitem}  
\usepackage[T1]{fontenc}
\usepackage{ae,aecompl}
\usepackage{newtxtext,newtxmath}
\usepackage{stackengine}
\usepackage{booktabs,makecell}
\usepackage{siunitx}
\usepackage{pifont}
\usepackage{tabularx}
\usepackage{cleveref}
\usepackage{multirow}
\usepackage{boldline}
\usepackage{array}
\usepackage{xfrac}
\usepackage{xcolor}
\usepackage{xspace}
\usepackage{orcidlink}
\usepackage{adjustbox}

\defcitealias{Etherington2022}{E22}

\newcommand\LDfull{Lensing\,\&\,Dynamics\xspace}

\newcommand\LD{\textrm{L\&D}\xspace}

\title[Beyond the bulge-halo conspiracy?]{Beyond the bulge-halo conspiracy? Density profiles of \\Early-type galaxies from extended-source strong lensing}
\author[Etherington et al.]{\parbox{\textwidth}{Amy Etherington\orcidlink{0000-0003-4642-7109}$^{1,2}$\thanks{Contact e-mail: \href{mailto:james.w.nightingale@durham.ac.uk}{james.w.nightingale@durham.ac.uk}}, 
James W.\ Nightingale\orcidlink{0000-0002-8987-7401}$^{1,2}$,
Richard Massey\orcidlink{0000-0002-6085-3780}$^{1,2}$, \\
Andrew Robertson\orcidlink{0000-0002-0086-0524}$^{3}$,
XiaoYue Cao$^{4,5}$, 
Aristeidis Amvrosiadis\orcidlink{0000-0002-4465-1564}$^{2}$, 
Shaun Cole\orcidlink{0000-0002-5954-7903}$^{2}$, \\
Carlos S.\ Frenk\orcidlink{0000-0002-2338-716X}$^{2}$,
Qiuhan He\orcidlink{0000-0003-3672-9365}$^{2}$,
David J. Lagattuta$^{1}$, 
Samuel Lange$^{2}$ \&
Ran Li$^{4,5}$ \\
}
\\
$^{1}$Department of Physics, Centre for Extragalactic Astronomy, Durham University, South Rd, Durham, DH1 3LE \\
$^{2}$Department of Physics, Institute for Computational Cosmology, Durham University, South Road, Durham DH1 3LE, UK \\
$^{3}$Jet Propulsion Laboratory, California Institute of Technology, 4800 Oak Grove Drive, Pasadena, CA 91109, USA\\
$^{4}$National Astronomical Observatories, Chinese Academy of Sciences, 20A Datun Road, Chaoyang District, Beijing 100012, China\\
$^{5}$School of Astronomy and Space Science, University of Chinese Academy of Sciences, Beijing 100049, China}

\date{}

\pubyear{2022}

\begin{document}
\label{firstpage}
\pagerange{\pageref{firstpage}--\pageref{lastpage}}
\maketitle

% Abstract of the paper
\begin{abstract}
%Richard's version
Observations suggest that the dark matter and stars in early-type galaxies `conspire' to produce a surprisingly simple distribution of total mass, $\rho(r)\propto\rho^{-\gamma}$, with $\gamma\approx2$. We measure the distribution of mass in 48 early-type galaxies that gravitationally lens a resolved background source. By fitting the source light in every pixel of images from the Hubble Space Telescope, we find a mean $\langle\gamma\rangle=2.075_{-0.024}^{+0.023}$ with intrinsic scatter between galaxies of $\sigma_\gamma=0.172^{+0.022}_{-0.032}$ for the overall sample. This is consistent with, and has similar precision to traditional techniques that employ spectroscopic observations to supplement lensing with mass estimates from stellar dynamics. Comparing measurements of $\gamma$ for individual lenses using both techniques, we find a statistically insignificant correlation of $-0.150^{+0.223}_{-0.217}$ between the two, indicating a lack of statistical power or deviations from a power-law density in certain lenses. At fixed surface mass density, we measure a redshift dependence, $\partial\langle\gamma\rangle/\partial z=0.345^{+0.322}_{-0.296}$, that is consistent with traditional techniques for the same sample of SLACS and GALLERY lenses. Interestingly, the consistency breaks down when we measure the dependence of $\gamma$ on the surface mass density of a lens galaxy. We argue that this is tentative evidence for an inflection point in the total-mass density profile at a few times the galaxy effective radius --- breaking the conspiracy.

\end{abstract}

% Select between one and six entries from the list of approved keywords.
% Don't make up new ones.

\subfile{1_Introduction.tex}
\subfile{2_Data.tex}   
\subfile{3_Method_Comp.tex}

\subfile{4_Correlations.tex}

\subfile{4_Redshift.tex}   
\subfile{5_Discussion.tex}

\subfile{6_Summary.tex}

\subfile{software.tex}

\section*{Acknowledgements}
AE is supported by STFC via grants ST/R504725/1 and ST/T506047/1.  
JN and RM are supported by STFC via grant ST/T002565/1, and the UK Space Agency via grant ST/W002612/1. XYC and RL acknowledge support from the National Nature Science Foundation of China (Nos.\ 11988101, 11773032, 12022306), science research grants from the China Manned Space Project (Nos.\ CMS-CSST-2021-B01, CMS-CSST-2021-A01) and support from the K.C.Wong Education Foundation. AA, SMC, CSF and QH acknowledge support from the European Research Council (ERC) through Advanced Investigator grant DMIDAS (GA 786910). This work used both the Cambridge Service for Data Driven Discovery (CSD3) and the DiRAC Data-Centric system, which are operated by the University of Cambridge and Durham University on behalf of the STFC DiRAC HPC Facility (www.dirac.ac.uk). These were funded by BIS capital grant ST/K00042X/1, STFC capital grants ST/P002307/1, ST/R002452/1, ST/H008519/1, ST/K00087X/1, STFC Operations grants ST/K003267/1, ST/K003267/1, and Durham University. DiRAC is part of the UK National E-Infrastructure.

\bibliographystyle{mnras}
\bibliography{references.bib, software.bib} % if your bibtex file is called example.bib

\appendix

\label{lastpage}

\end{document}

%% file: 1_Introduction.tex
\section{Introduction}\label{Introduction}

Early-type galaxies (hereafter ETGs) are the end product of the hierarchical merging paradigm central to the $\Lambda$-Cold Dark Matter (CDM) cosmological model \citep{White1978, Cole1994}. They are built from the successive mergers between more and more massive objects, and hence provide tests of the entire process of galaxy formation and evolution. The distribution of mass in their baryon-dominated inner regions are especially sensitive, because baryonic physics significantly redistributes mass at various stages of evolution. The inner mass-density profile may become steeper as a result of higher baryon densities from dissipative gas cooling processes and the inflow of gas \citep{Blumenthal1986, Silk1993, Velliscig2014}. They may become softened by outflows of gas driven by feedback processes such as active galactic nuclei and supernovae \citep{Velliscig2014, Dubois2013}. Measurements of ETG inner mass-density profiles are therefore fundamental in understanding the relative strength and timing of these physical processes.

Observations have shown that the mean distribution of dark plus baryonic matter in the central few effective radii of ETGs is such that their combined density profile is roughly isothermal, $\rho(r) \propto r^{-\gamma}$, with $\gamma\sim2$. This has been consistently observed by many observational techniques: dynamically modelled local ETGs \citep{Tortora2014a, Serra2016, Li2019, Cappellari2013}, X-ray studies \citep{Humphrey2006, Humphrey2010}, weak lensing \citep{Gavazzi2007}, and combined strong lensing and dynamical modelling \citep{Koopmans2009, Auger2010, Li2018}. The latter is the most prevalent of these results, with the `standard' procedure developed by \cite{Treu2002} constraining the total mass inside two different radii: the galaxy light's effective radius from measurements of the velocity dispersion, and the galaxy mass's Einstein radius from lensing. In this way, \cite{Auger2010} measured a mean logarithmic density slope $\langle \gamma\rangle = 2.078\pm 0.027$, with intrinsic scatter between galaxies of $\sigma_\gamma$ = $0.16\pm0.02$, for the largest single sample of strong lenses that make up the Sloan Lens ACS (SLACS) survey \citep{Bolton2008a}.

The near-isothermality of mass in ETGs is often termed the `bulge-halo conspiracy', referring to the apparent coincidence that despite diverse assembly histories, and although neither their baryonic nor dark matter components follow a single power law, their sum approximately does \citep{Treu2006, Humphrey2010}. The galaxies' homogeneity is further evident in the well-known ETG scaling laws such as the fundamental plane relations \citep{Djorgovski1987} and the $M_{\rm BH}-\sigma_{\rm c}$ relation \citep{Hyde2009}. Furthermore, the total mass-density slopes correlate with a number of galaxy parameters including effective radius, stellar surface mass density, and central dark matter fraction, as well as being observed to mildly soften with increasing redshift up to $z\sim 1.0$ \citep{Auger2010, Ruff2011, Sonnenfeld2013a, Li2018}.

Numerical simulations are invaluable in understanding the origin of these empirical relations, and are now beginning to account for the physical processes involved in their formation. The current consensus for the formation of ETGs, often referred to as a `two phase' assembly \citep{Oser2010}, begins with an initial stage of active star formation and adiabatic contraction at redshift $z\gtrsim2$, followed by growth through major and minor merging events to the present \citep{Naab2009, VanDokkum2010, Remus2017}. However, details of the physical processes that modify the mass distributions throughout this formation process are yet to be well understood. Fine-tuning between the baryonic and dark matter distributions would be necessary to produce the distribution of near-isothermal total mass profiles that are observed, a result hydrodynamic simulations have been unable to accomplish whilst simultaneously reproducing the observed distribution of dark matter fractions \citep{Duffy2010, Dubois2013, Xu2017}. It is unclear whether this discrepancy is a result of an inadequacy in the cosmological simulations or a systematic bias in the determination of the observed mass-density slopes. 

Comparing observed and simulated mass-density slopes is difficult. \cite{wang2020} demonstrated that IllustrisTNG reproduces many of the observed mass-density slope correlations, assuming the best fit total power-law density slope within the radial interval [0.4R$_{1/2}$, 4R$_{1/2}$] of their simulated sample of ETGs (see also \citealt{Mukherjee2018, Mukherjee2021, Peirani2019}). Although tensions do exist, the authors find a negative correlation with central velocity dispersion ($\sigma$) for the simulated galaxies whereas observational datasets tend towards a positive correlation. This is the case for both strong lensing and dynamical observations. \cite{Li2019} show that  both IllustrisTNG and EAGLE simulations are unable to reproduce the $\gamma-\sigma$ trend observed from a dynamical analysis of over 2000 galaxies in the SDSS-IV (Sloan Digital Sky Survey IV) MaNGA survey; both simulations typically predict shallower slopes than those observed for the high velocity dispersion galaxies in their sample. 

Furthermore, cosmological simulations typically exhibit a mild steepening of the density slope with redshift up to $z\sim2.0$, in contrast with the mild softening observed \citep{Johansson2012, Remus2017, Xu2017}. \cite{Xu2017} and \cite{Remus2017} demonstrated the Illustris and Magneticum simulations show better agreement with the observations when using a different estimator for the power-law slope that better resembles the observational methods. However, the estimator differs between the two studies, and a direct comparison to observations must still be approached with caution. For example, \cite{Xu2017} note that their observational slope estimator results in a sampling bias whereby the simulated sample have relatively lower mean slopes due to a larger fraction of systems with lower mass and/or smaller normalised Einstein radii. An appropriate comparison would require strictly adopting observational criteria to estimate the slopes and select the simulated samples. Further to this, the necessary observational data for a large sample of galaxies out to high redshifts would allow for a more complete comparison to the simulations.

In this work, we build on the results presented in \citet[][hereafter E22]{Etherington2022}, who used strong gravitational lensing \textit{alone} to measure the total mass-density profiles of 59 lenses from the SLACS \citep{Bolton2006} and BELLS GALLERY \citep{Shu2016a} samples. Previous lensing and dynamical analyses exploited only one lensing observable, the Einstein radius, and inferred the logarithmic density slope by combining this with measurements of stellar kinematics. However, E22 used the fact that light rays emitted from opposite sides of an extended source are deflected by different amounts. Using the lens modelling software \texttt{PyAutoLens} \citep{Nightingale2018, Nightingale2021}, \citetalias{Etherington2022} fitted the full surface brightness profile of observed arcs to constrain the total mass-density profile of these 59 lens galaxies. The measurement was automated, to ensure that it will also be able to exploit the tens of thousands of lenses expected to be observed by LSST and Euclid \citep{Collett2015}. Since this measurement uses only imaging data, and does not require expensive spectroscopy for stellar kinematics, it has the potential to measure the formation and accumulation of mass around galaxies out to redshift $z = 2.0$ and beyond \citep{Sonnenfeld2021, Sonnenfeld2021a, Sonnenfeld2022}.

We introduce the samples of lenses that we study in this work which have total mass density slopes derived from either lensing-only, lensing and dynamics, or both, in Section~\ref{Data}. In Section~\ref{method comp} we investigate the assumption of the power law mass distribution by comparing to what extent the two methods infer the same slope. We then quantify the dependence of the slopes, measured with both methods, on redshift in Section~\ref{multiple covariates}, before discussing the results in Section ~\ref{Discussion}, and concluding in Section~\ref{Summary}. Throughout this work we assume a Planck 2015 cosmological model \citep{Ade2016}.

%% file: 2_Data.tex
\section{Observational Samples of Galaxies}\label{Data}

\subsection{Complete sample: Lensing \& Dynamics (L\&D) measurements from the literature}\label{complete sample}

\begin{figure}
    \centering
    \includegraphics{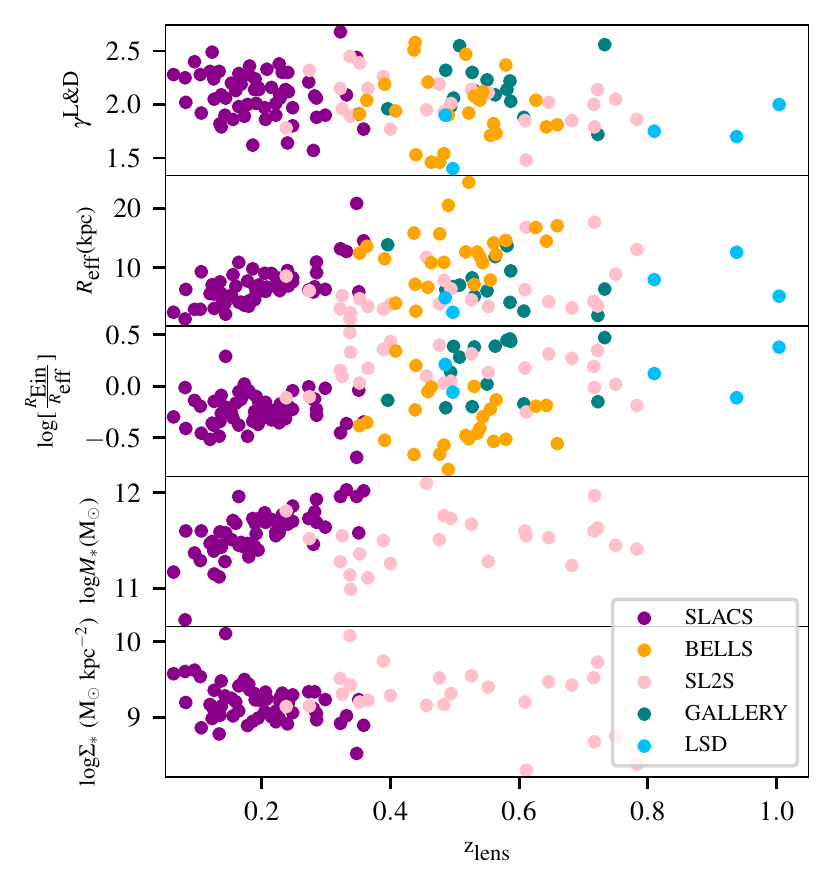}
    \caption{Various galaxy quantities plotted as a function of redshift for the complete lensing\&dynamics sample. Stellar masses (hence stellar surface mass densities) have not been measured for BELLS, GALLERY, or LSD samples. Where necessary throughout this study we instead use total masses, derived from equation~\ref{eq: M total}.}
    \label{Figure: LD galaxy observables}
\end{figure}

Hundreds of galaxy-scale strong lenses have been discovered by dedicated surveys during the past two decades, with measurements of their mass profiles by e.g.\ \cite{Treu2006, Koopmans2006, Auger2010, Sonnenfeld2013a} and \cite{Li2018}. The method, initially developed by \cite{Treu2002}, models the stellar plus dark matter distribution of total mass in each galaxy as $\rho\propto r^{-\gamma}$. By further assuming a stellar density profile (treated as a massless tracer of the total density profile), with effective radii fixed to those observed (typically from de Vaucouleurs models), the spherical Jeans equations can be solved to calculate the velocity dispersion for a given model. The total mass-density slope $\gamma$ can then be constrained using the mass within the Einstein radius and the stellar velocity dispersion by comparing the model values to those observed -- the Einstein radius is typically measured from fits to imaging data assuming a singular isothermal ellipse mass model \citep[SIE;][]{Kormann1994}. 

We have collated a ``complete \LDfull sample'' of 123 lens galaxies from Sloan Lens ACS (SLACS), \citep{Bolton2006, Auger2010}, BOSS Emission Line Lens (BELLS) \citep{Brownstein2012}, BELLS GALaxy-Ly$\alpha$ EmitteR sYstems (GALLERY) \citep{Shu2016a, Shu2016b}, Strong Lensing Legacy (SL2S) \citep{Gavazzi2012} surveys, and Lenses Structure and Dynamics  (LSD) surveys \cite{Treu2004} for which measurements of the total-mass density slope from the combined lensing and dynamics (L\&D) analysis have previously been carried out. Lens galaxies were selected in the following different ways in the various surveys:
\begin{itemize}
    \item \textbf{SLACS (50 lenses)}: spectroscopic search within the SDSS database, using a 3$\arcsec$ fibre, examining residual spectra for higher redshift emission lines\footnote{\cite{Auger2010} find that six of the SLACS galaxies are significant outliers of the fundamental hyper-plane relation (the relationship between the effective radius, velocity dispersion, central stellar mass, and central total mass), which may be a a result of significantly underestimated velocity dispersion errors \cite{Jiang2007}. In keeping with previous studies we remove those from our sample.}.
    \item \textbf{BELLS (25 lenses)}: spectroscopic search within the BOSS database, using a 2$\arcsec$ fibre, examining residual spectra for higher redshift emission lines.
    \item \textbf{GALLERY (15 lenses)}: same technique as BELLS with an additional selection for higher redshift, compact Lyman-$\alpha$-emitting (LAE) source galaxies.
    \item \textbf{SL2S (25 lenses)}: imaging data is analysed for an excesses of blue features that indicate the presence of lensed features \citep{Gavazzi2014}.
    \item \textbf{LSD (5 lenses)}: systems selected from  the CfA-Arizona Space Telescope Lens Survey (CASTLeS)\footnote{see the CASTLeS web-page at \url{http://cfa-www.harvard.edu/castles/}} sample of known galaxy-scale systems for their morphology (E/S0) and brightness (I $\lesssim$ 22).  
\end{itemize}
To our knowledge, this is the first time all these observations have been studied in one analysis. As well as L\&D total mass density slopes, we gather literature measurements of a number of galaxy observables including velocity dispersions, effective radii, Einstein radii (which we normalise by the effective radii throughout this study), stellar masses, and stellar surface mass densities $(\Sigma^*=M^*/(2\pi\textrm{R}_\textrm{eff}^2))$, which are plotted as a function of redshift of the lens galaxy in Figure~\ref{Figure: LD galaxy observables}.

Total-mass density slopes have been shown to correlate with both total and stellar surface mass-densities \citep{Auger2010a, Sonnenfeld2013a}. We must account for this relationship if we wish to study how the density profile depends on redshift, because stellar density also evolves with redshift. Notably, stellar masses (hence stellar surface mass densities) have not been measured for BELLS, GALLERY, or LSD samples. Following \citet{Auger2010}\footnote{the convention for $\Sigma_\textrm{tot}$ in \citet{Auger2010} does not include a $\pi$ in the denominator, whereas the stellar surface mass density $\Sigma^*$ given by \citet{Sonnenfeld2013a} does. We follow the convention of each paper that a quantity is taken from and therefore retain $\pi$ in the denominator for $\Sigma^*$ but drop it for $\Sigma_\textrm{tot}$.}, we therefore calculate \textit{total} surface mass densities
\begin{equation}\label{eq: sigma tot}
    \Sigma_\textrm{tot} = \frac{M_\textrm{tot}}{R_\textrm{eff}^2},
\end{equation}
where $R_\textrm{eff}$ is the effective radius of the galaxy and 
\begin{equation}\label{eq: M total}
    M_\textrm{tot}=\Sigma_\textrm{crit}\pi R_\textrm{Ein}^{\gamma-1}\bigg(\frac{R_\textrm{eff}}{2}\bigg)^{3-\gamma}
\end{equation}
is the total projected mass within half the effective radius inferred from power-law models with Einstein radii $R_\textrm{Ein}$ and total mass-density slope $\gamma$. The total projected mass is calculated within half the effective radius, which typically resembles closely the Einstein radius, to reduce errors from extrapolating the power law model. 

\subsection{Complete sample: new measurements using Lensing-only}\label{complete lensing samples}
\begin{figure}
    \centering
    \includegraphics[width=\linewidth]{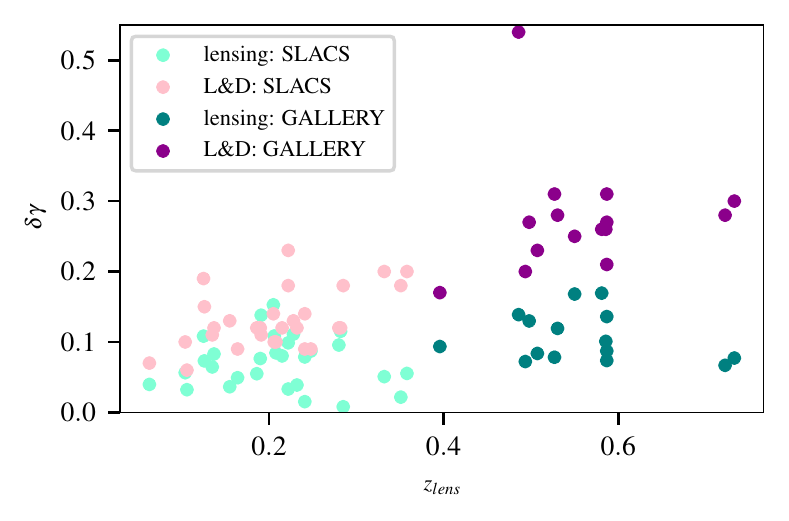}
    \caption{Measurement uncertainties on the slopes from lensing only and lensing \& dynamics as a function of redshift of the lens galaxy.}
    \label{Figure: err v redshift}
\end{figure}

If a lensed galaxy is spatially resolved, the apparent shape of the arc can be used to infer the distribution of total mass-density around a foreground lens, without any spectroscopic information about stellar kinematics. The source flux in every image pixel can be ray-traced back to the source plane, and the shape of the source galaxy is modelled as a sum of analytic functions \citep{Tessore2016a}, possibly combined with a basis of shapelets \citep{Birrer2015, Shajib2018}, or a pixelised source \citep{Warren2003, Suyu2006, Dye2005, Vegetti2009, Nightingale2015, Nightingale2018, Joseph2019, Galan2021}. The configuration of ray-tracing required to map multiple images in the lens plane onto consistent morphologies in the source plane constrains parameters of the mass model, including its logarithmic density slope $\gamma$.

\citetalias{Etherington2022} used this approach to model a sub-sample of 43 SLACS and 15 GALLERY lenses. Here we consider only the 53 ``Gold'' and 4 ``Silver'' lenses for which an automated analysis reliably fitted the data without residuals \citepalias[see][for the detail of the categories]{Etherington2022}. We refer to this sample of 57 lenses as the ``complete lensing-only sample''. 
We note that, with the lensing-only technique, the density profile constraints from compact LAE sources in the GALLERY sample are not as tight as constraints from the more extended sources in the SLACS sample. However, the slopes of GALLERY lenses are still better constrained than the slopes measured for the same lenses using the \LD analysis (Figure~\ref{Figure: err v redshift}).

\subsection{Overlapping sample: galaxies with both Lensing-only and Lensing \& Dynamics measurements}\label{overlapping samples}
\begin{figure}
    \centering
    \includegraphics{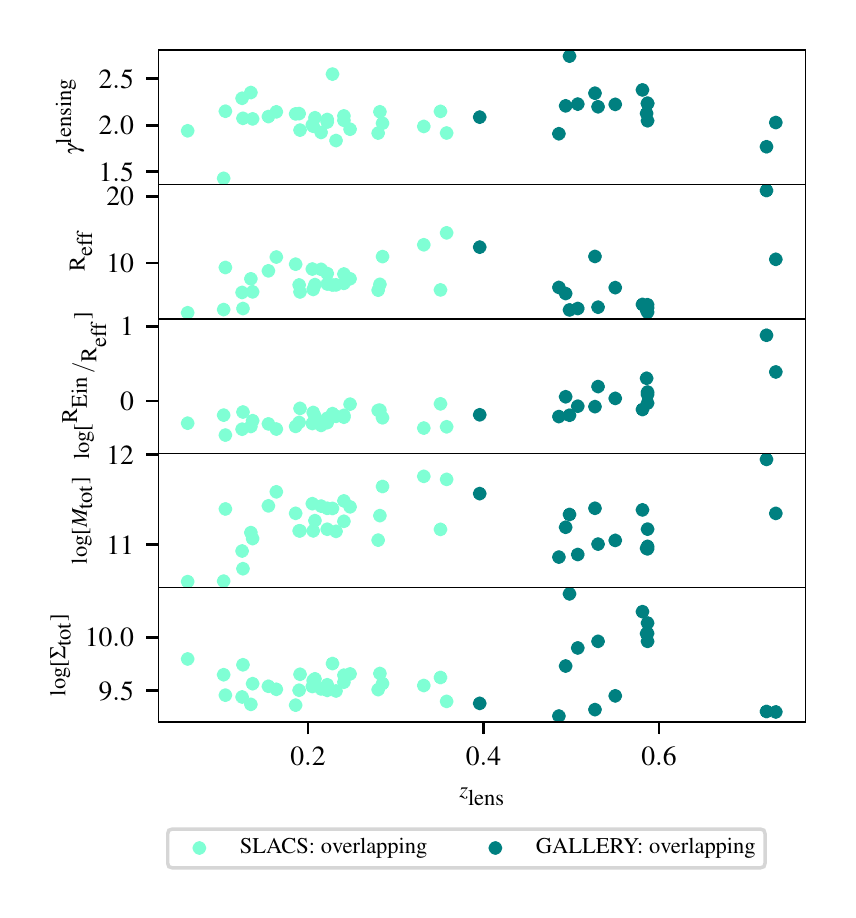}
    \caption{Galaxy observables as a function of lens redshift for the overlapping sample. From top to bottom panel the quantities are as follows: lensing only total mass density slope ($\gamma^\textrm{lensing}$), de Vaucouleurs effective radii $R_\textrm{eff}$ in units of kpc, Einstein radius normalised by the effective radius (log[$^{\textrm{R}_\textrm{Ein}}/_{\textrm{R}_\textrm{eff}}$]), total mass within half the effective radius (log[$M_\textrm{tot}$]), and total surface mass density (log[$\Sigma_\textrm{tot}$]). The total masses and total surface mass densities plotted here are those derived from the lensing quantities ($\gamma^\textrm{lensing}$ and $R_\textrm{Ein}^\textrm{PL}$ in equations~\ref{eq: M total} and ~\ref{eq: sigma tot}), we note that they do not change significantly when derived from L\&D quantities (see Table~\ref{table: lens parameters} for both values).}
    \label{Figure: overlapping sample observables v redshift}
\end{figure}

To directly compare the two methods, we select the subset of lenses whose density slope has been measured by both lensing-only {\it and} \LD. %For a direct comparison between the two methods, we select the lenses that have measurements of the density slope from both the lensing-only and \LD analyses. 
This requires excluding 1 GALLERY and 6 SLACS lenses from the complete lensing-only sample whose mass slopes have not previously been measured using the L\&D method. As in the complete sample, we also exclude 3 SLACS lenses suspected to have anomalous measurements of velocity dispersion. We shall refer to the remaining 48 lenses as the ``overlapping sample''. 

As for the complete \LD sample we gather literature measurements of a number of galaxy observables including velocity dispersions, effective radii, and normalised Einstein radii (plotted as a function of redshift in Figure~\ref{Figure: overlapping sample observables v redshift}). We also calculate total masses (equation~\ref{eq: M total}) and surface mass densities (equation~\ref{eq: sigma tot}). Note that the effective radius of all galaxies in the overlapping sample has been measured at least twice: assuming de Vaucouleurs surface brightness profiles in L\&D analyses \citep[e.g.][]{Auger2010} and double S\'ersic profiles in the lensing analysis \citetalias{Etherington2022} (see Table~\ref{table: lens parameters}). Since the L\&D mass density slopes were calculated using de Vaucouleurs effective radii, we use these for consistency with previous literature whenever we quote an effective radius. The lensing-only analyses do not use their measurements of effective radius. A mildly positive trend of $R_{\rm eff}$ with $z_{\rm lens}$ is seen, which is reported by other studies (e.g. \citealt{Sonnenfeld2013a}) and related to how $R_{\rm eff}$ correlates with mass.

{\renewcommand{\arraystretch}{1.5}
\begin{table*}
\centering
\begin{adjustbox}{max width=\textwidth}
\begin{tabular}{llrrrllrlrrrrr}
\toprule
\multicolumn{1}{p{0.7cm}}{\centering Sample} & 
\multicolumn{1}{p{1.3cm}}{\centering Lens Name} &      
\multicolumn{1}{p{0.8cm}}{\centering $z_\textrm{lens}$} & 
\multicolumn{1}{p{0.8cm}}{\centering $z_\textrm{source}$} & 
\multicolumn{1}{p{1.6cm}}{\centering $\sigma$ $(\mathrm{km}\,\mathrm{s}^{-1})$} &    
\multicolumn{1}{p{1.2cm}}{\centering $R_\textrm{Ein}^\textrm{PL}$ $(\arcsec)$} & 
\multicolumn{1}{p{1.2cm}}{\centering $R_\textrm{Ein}^\textrm{SIE}$ $(\arcsec)$} & 
\multicolumn{1}{p{1cm}}{\centering $\gamma^\textrm{lensing}$} & 
\multicolumn{1}{p{1cm}}{\centering $\gamma^\textrm{L\&D}$} & 
\multicolumn{1}{p{1.2cm}}{\centering $R_\textrm{eff}^\textrm{dev}$ (kpc)} & 
\multicolumn{1}{p{1.5cm}}{\centering $\log [\frac{M_\textrm{tot}^\textrm{lensing}}{M_{\rm \odot}}]$} & 
\multicolumn{1}{p{1.5cm}}{\centering $\log [\frac{M_\textrm{tot}^\textrm{L\&D}}{M_{\rm \odot}}]$} & 
\multicolumn{1}{p{1.5cm}}{\centering $\log [\frac{\Sigma_\textrm{tot}^\textrm{lensing}}{M_{\rm \odot}}]$} & 
\multicolumn{1}{p{1.5cm}}{\centering $\log [\frac{\Sigma_\textrm{tot}^\textrm{L\&D}}{M_{\rm \odot}}]$} \\
\midrule
\multirow{29}{*}{SLACS} & J0216-0813 &      0.332 &        0.523 &   $319\pm23$ &  $1.183^\textrm{+0.014}_\textrm{-0.011}$ &             1.16 &  $1.99^\textrm{+0.05}_\textrm{-0.06}$ &        $2.09\pm0.2$ &  12.74 &              11.76 &          11.75 &               9.55 &           9.53 \\
        & J0252+0039 &      0.280 &        0.982 &   $170\pm12$ &  $1.024^\textrm{+0.004}_\textrm{-0.002}$ &             1.04 &  $1.92^\textrm{+0.08}_\textrm{-0.11}$ &        $1.57\pm0.12$ &  5.90 &              11.05 &          11.00 &               9.51 &           9.46 \\
        & J0330-0020 &      0.351 &        1.071 &   $213\pm21$ &  $1.088^\textrm{+0.009}_\textrm{-0.012}$ &             1.10 &  $2.15^\textrm{+0.02}_\textrm{-0.02}$ &        $1.91\pm0.18$ &  5.94 &              11.17 &          11.11 &               9.62 &           9.57 \\
        & J0728+3835 &      0.206 &        0.688 &   $210\pm11$ &  $1.244^\textrm{+0.012}_\textrm{-0.008}$ &             1.25 &   $1.99^\textrm{+0.12}_\textrm{-0.1}$ &        $1.86\pm0.1$ & 6.01 &              11.16 &          11.14 &               9.60 &           9.58 \\
        & J0822+2652 &      0.241 &        0.594 &   $254\pm15$ &  $1.129^\textrm{+0.011}_\textrm{-0.018}$ &             1.17 &   $2.1^\textrm{+0.08}_\textrm{-0.07}$ &        $2.12\pm0.14$ & 6.93 &              11.26 &          11.28 &               9.58 &           9.60 \\
        & J0912+0029 &      0.164 &        0.324 &   $304\pm16$ &  $1.393^\textrm{+0.011}_\textrm{-0.007}$ &             1.63 &  $2.14^\textrm{+0.05}_\textrm{-0.05}$ &        $1.98\pm0.09$ & 10.89 &              11.59 &          11.68 &               9.51 &           9.60 \\
        & J0936+0913 &      0.190 &        0.588 &   $236\pm12$ &  $1.081^\textrm{+0.004}_\textrm{-0.005}$ &             1.09 &  $2.13^\textrm{+0.08}_\textrm{-0.08}$ &        $2.24\pm0.12$ &  6.69 &              11.16 &          11.16 &               9.51 &           9.51 \\
        & J0946+1006 &      0.222 &        0.609 &   $253\pm21$ &  $1.409^\textrm{+0.001}_\textrm{-0.001}$ &             1.38 &  $2.06^\textrm{+0.03}_\textrm{-0.03}$ &        $2.01\pm0.18$ &  8.41 &              11.41 &          11.39 &               9.56 &           9.54 \\
        & J0956+5100 &      0.241 &        0.470 &   $323\pm17$ &  $1.314^\textrm{+0.002}_\textrm{-0.001}$ &             1.33 &  $2.05^\textrm{+0.02}_\textrm{-0.02}$ &        $2.3\pm0.09$ & 8.33 &              11.49 &          11.51 &               9.65 &           9.67 \\
        & J0959+0410 &      0.126 &        0.535 &   $196\pm13$ &  $0.985^\textrm{+0.014}_\textrm{-0.017}$ &             0.99 &  $2.08^\textrm{+0.07}_\textrm{-0.07}$ &        $2.05\pm0.15$ &  3.14 &              10.74 &          10.73 &               9.74 &           9.74 \\
        & J1020+1122 &      0.282 &        0.553 &   $279\pm18$ &  $1.065^\textrm{+0.011}_\textrm{-0.009}$ &             1.20 &  $2.15^\textrm{+0.11}_\textrm{-0.12}$ &        $2.08\pm0.12$ &  6.78 &              11.32 &          11.37 &               9.66 &           9.71 \\
        & J1023+4230 &      0.191 &        0.696 &   $238\pm15$ &  $1.411^\textrm{+0.009}_\textrm{-0.009}$ &             1.41 &  $1.95^\textrm{+0.16}_\textrm{-0.12}$ &        $2.01\pm0.11$ &  5.63 &              11.16 &          11.17 &               9.65 &           9.67 \\
        & J1029+0420 &      0.104 &        0.615 &   $208\pm11$ &    $0.947^\textrm{+0.01}_\textrm{-0.01}$ &             1.01 &  $1.43^\textrm{+0.05}_\textrm{-0.06}$ &        $2.28\pm0.1$ &  2.98 &              10.60 &          10.71 &               9.65 &           9.76 \\
        & J1142+1001 &      0.222 &        0.504 &   $216\pm22$ &  $0.908^\textrm{+0.024}_\textrm{-0.027}$ &             0.98 &    $2.03^\textrm{+0.1}_\textrm{-0.1}$ &        $1.9\pm0.23$  &  6.83 &              11.17 &          11.21 &               9.50 &           9.54 \\
        & J1143-0144 &      0.106 &        0.402 &   $247\pm13$ &  $1.611^\textrm{+0.013}_\textrm{-0.014}$ &             1.68 &  $2.15^\textrm{+0.03}_\textrm{-0.03}$ &       $1.92\pm0.06$ & 9.32 &              11.40 &          11.45 &               9.46 &           9.52 \\
        & J1205+4910 &      0.215 &        0.481 &   $269\pm14$ &  $1.218^\textrm{+0.008}_\textrm{-0.008}$ &             1.22 &  $1.92^\textrm{+0.07}_\textrm{-0.09}$ &        $2.16\pm0.12$  &   9.04 &              11.43 &          11.42 &               9.52 &           9.51 \\
        & J1218+0830 &      0.135 &        0.717 &   $207\pm11$ &   $1.217^\textrm{+0.01}_\textrm{-0.008}$ &             1.45 &  $2.35^\textrm{+0.07}_\textrm{-0.06}$ &        $1.82\pm0.11$  & 7.61 &              11.14 &          11.26 &               9.37 &           9.50 \\
        & J1250+0523 &      0.232 &        0.795 &   $247\pm14$ &  $1.144^\textrm{+0.006}_\textrm{-0.005}$ &             1.13 &  $1.84^\textrm{+0.04}_\textrm{-0.04}$ &       $2.3\pm0.12$ &  6.69 &              11.15 &          11.19 &               9.50 &           9.54 \\
        & J1402+6321 &      0.205 &        0.481 &   $255\pm17$ &  $1.349^\textrm{+0.005}_\textrm{-0.007}$ &             1.35 &   $2.0^\textrm{+0.18}_\textrm{-0.13}$ &        $1.97\pm0.14$ &  9.08 &              11.46 &          11.46 &               9.54 &           9.54 \\
        & J1420+6019 &      0.063 &        0.535 &   $199\pm10$ &  $1.075^\textrm{+0.002}_\textrm{-0.002}$ &             1.04 &  $1.94^\textrm{+0.04}_\textrm{-0.04}$ &       $2.28\pm0.07$ &  2.50 &              10.59 &          10.58 &               9.80 &           9.78 \\
        & J1430+4105 &      0.285 &        0.575 &   $309\pm32$ &  $1.481^\textrm{+0.002}_\textrm{-0.002}$ &             1.52 &  $2.02^\textrm{+0.01}_\textrm{-0.01}$ &        $2.06\pm0.18$ &  10.96 &              11.65 &          11.66 &               9.57 &           9.58 \\
        & J1451-0239 &      0.125 &        0.520 &   $214\pm14$ &   $0.96^\textrm{+0.017}_\textrm{-0.015}$ &             1.04 &   $2.29^\textrm{+0.1}_\textrm{-0.11}$ &        $2.24\pm0.19$  &  5.56 &              10.93 &          10.98 &               9.44 &           9.49 \\
        & J1525+3327 &      0.358 &        0.717 &   $150\pm10$ &   $1.29^\textrm{+0.012}_\textrm{-0.007}$ &             1.31 &  $1.92^\textrm{+0.06}_\textrm{-0.05}$ &        $1.77\pm0.2$ &   14.54 &              11.73 &          11.74 &               9.40 &           9.41 \\
        & J1627-0053 &      0.208 &        0.524 &   $283\pm15$ &  $1.217^\textrm{+0.002}_\textrm{-0.002}$ &             1.23 &  $2.08^\textrm{+0.08}_\textrm{-0.09}$ &       $2.33\pm0.1$ &  6.74 &              11.27 &          11.30 &               9.61 &           9.64 \\
        & J1630+4520 &      0.248 &        0.793 &   $269\pm16$ &  $1.791^\textrm{+0.006}_\textrm{-0.004}$ &             1.78 &  $1.96^\textrm{+0.09}_\textrm{-0.08}$ &        $1.77\pm0.2$ &     7.62 &              11.42 &          11.42 &               9.66 &           9.66 \\
        & J2238-0754 &      0.137 &        0.713 &   $191\pm11$ &  $1.268^\textrm{+0.004}_\textrm{-0.003}$ &             1.27 &  $2.07^\textrm{+0.09}_\textrm{-0.07}$ &        $1.79\pm0.12$ &    5.65 &              11.07 &          11.06 &               9.57 &           9.56 \\
        & J2300+0022 &      0.228 &        0.463 &   $273\pm17$ &  $1.219^\textrm{+0.008}_\textrm{-0.005}$ &             1.24 &  $2.55^\textrm{+0.07}_\textrm{-0.16}$ &        $2.06\pm0.13$ &    6.68 &              11.40 &          11.35 &               9.75 &           9.70 \\
        & J2303+1422 &      0.155 &        0.517 &    $240\pm6$ &  $1.628^\textrm{+0.007}_\textrm{-0.005}$ &             1.62 &  $2.09^\textrm{+0.04}_\textrm{-0.04}$ &        $1.86\pm0.13$ &   8.81 &              11.43 &          11.43 &               9.54 &           9.54 \\
        & J2341+0000 &      0.186 &        0.807 &   $196\pm13$ &  $1.338^\textrm{+0.009}_\textrm{-0.005}$ &             1.44 &  $2.12^\textrm{+0.06}_\textrm{-0.05}$ &       $1.62\pm0.12$ &    9.81 &              11.35 &          11.41 &               9.37 &           9.42 \\
\cline{1-14}
\multirow{15}{*}{GALLERY} & J0029+2544 &      0.587 &        2.450 &   $240\pm45$ &  $1.347^\textrm{+0.014}_\textrm{-0.012}$ &             1.34 &  $2.05^\textrm{+0.12}_\textrm{-0.15}$ &        $2.03\pm0.27$ &    9.46 &              11.43 &          11.42 &               9.47 &           9.47 \\
        & J0201+3228 &      0.396 &        2.821 &   $245\pm20$ &  $1.713^\textrm{+0.011}_\textrm{-0.005}$ &             1.70 &   $2.09^\textrm{+0.09}_\textrm{-0.1}$ &   $1.96\pm0.17$ &     13.88 &              11.61 &          11.59 &               9.33 &           9.31 \\
        & J0237-0641 &      0.486 &        2.249 &   $295\pm89$ &   $0.619^\textrm{+0.02}_\textrm{-0.025}$ &             0.65 &   $1.91^\textrm{+0.18}_\textrm{-0.1}$ &  $2.32\pm0.54$ &   6.31 &              10.86 &          10.92 &               9.26 &           9.32 \\
        & J0742+3341 &      0.494 &        2.363 &   $221\pm28$ &   $1.241^\textrm{+0.01}_\textrm{-0.013}$ &             1.22 &  $2.21^\textrm{+0.06}_\textrm{-0.08}$ &   $1.98\pm0.2$ &    6.49 &              11.26 &          11.17 &               9.63 &           9.54 \\
        & J0755+3445 &      0.722 &        2.635 &   $302\pm52$ &  $2.073^\textrm{+0.005}_\textrm{-0.004}$ &             2.05 &  $1.77^\textrm{+0.08}_\textrm{-0.05}$ &  $1.72\pm0.28$ &   1.95 &              10.68 &          10.62 &              10.10 &          10.04 \\
        & J0856+2010 &      0.507 &        2.234 &   $337\pm54$ &   $0.951^\textrm{+0.035}_\textrm{-0.04}$ &             0.98 &  $2.23^\textrm{+0.08}_\textrm{-0.09}$ &  $2.55\pm0.23$ &    7.07 &              11.17 &          11.26 &               9.47 &           9.56 \\
        & J0918+5105 &      0.581 &        2.403 &   $289\pm49$ &  $1.645^\textrm{+0.005}_\textrm{-0.009}$ &             1.60 &  $2.38^\textrm{+0.16}_\textrm{-0.18}$ &   $2.14\pm0.26$  &   13.70 &              11.74 &          11.67 &               9.46 &           9.40 \\
        & J1110+2808 &      0.733 &        2.400 &   $207\pm39$ &  $0.904^\textrm{+0.027}_\textrm{-0.026}$ &             0.98 &  $2.03^\textrm{+0.09}_\textrm{-0.07}$ &  $1.88\pm0.3$ &     2.91 &              10.81 &          10.74 &               9.88 &           9.81 \\
        & J1110+3649 &      0.587 &        2.502 &  $546\pm165$ &  $1.151^\textrm{+0.001}_\textrm{-0.001}$ &             1.16 &  $2.23^\textrm{+0.07}_\textrm{-0.08}$ &  $2.56\pm0.32$ &    5.82 &              11.23 &          11.37 &               9.70 &           9.84 \\
        & J1116+0915 &      0.550 &        2.454 &   $280\pm55$ &  $0.811^\textrm{+0.053}_\textrm{-0.054}$ &             1.03 &  $2.22^\textrm{+0.16}_\textrm{-0.17}$ &   $2.23\pm0.25$  &   6.09 &              11.04 &          11.17 &               9.47 &           9.60 \\
        & J1141+2216 &      0.586 &        2.762 &   $299\pm44$ &  $1.283^\textrm{+0.027}_\textrm{-0.019}$ &             1.27 &  $2.13^\textrm{+0.09}_\textrm{-0.11}$ &    $2.22\pm0.26$ &    4.16 &              11.10 &          11.15 &               9.86 &           9.91 \\
        & J1201+4743 &      0.498 &        2.126 &   $234\pm43$ &  $1.171^\textrm{+0.004}_\textrm{-0.002}$ &             1.18 &  $2.74^\textrm{+0.05}_\textrm{-0.21}$ &      $2.09\pm0.27$ &   11.15 &              11.49 &          11.42 &               9.39 &           9.33 \\
        & J1226+5457 &      0.587 &        2.732 &   $251\pm26$ &  $1.398^\textrm{+0.004}_\textrm{-0.003}$ &             1.37 &   $2.24^\textrm{+0.07}_\textrm{-0.1}$ &    $2.06\pm0.21$ &     7.41 &              11.40 &          11.32 &               9.66 &           9.58 \\
        & J2228+1205 &      0.530 &        2.832 &   $263\pm50$ &   $1.21^\textrm{+0.024}_\textrm{-0.024}$ &             1.28 &    $2.2^\textrm{+0.14}_\textrm{-0.1}$ &     $2.13\pm0.28$  &   5.16 &              11.16 &          11.16 &               9.73 &           9.73 \\
        & J2342-0120 &      0.527 &        2.265 &   $274\pm43$ &  $1.091^\textrm{+0.006}_\textrm{-0.004}$ &             1.11 &  $2.34^\textrm{+0.07}_\textrm{-0.09}$ &    $2.3\pm0.31$  &   8.28 &              11.33 &          11.33 &               9.49 &           9.49 \\
\bottomrule
\end{tabular}
\end{adjustbox}
    \caption{Lens parameters for the overlapping sample. Errors are quoted at $1\sigma$ confidence intervals.} 
    \label{table: lens parameters}
\end{table*}}

%% file: 3_Method_Comp.tex
\section{Do the lensing-only and lensing \& dynamics methods measure the same density slopes?}\label{method comp}

Although the lensing-only and \LD methods aim to measure the same quantity $\gamma$, the assumption of the power-law profile is critical in this endeavour. The \LD analysis is averaged between mass measurements at the Einstein and effective radii, whereas the lensing method is constrained by the pixel information of the source galaxy that, by definition, occurs near the Einstein radius. Any deviation of the galaxy's true mass-density profile from a power law could therefore lead to biases on $\gamma$ that behave differently between the two methods (e.g.\ \citealt{Schneider2013, Kochanek2020, Cao2021}). We therefore investigate to what extent the methods infer the same slope, first by comparing the sample averages in Section~\ref{section: sample comparison}, measurements of individual galaxies in Section~\ref{section: individual comparison}, then correlations between galaxies' slopes and other observable quantities in Section~\ref{section: correlations}. 

\subsection{Comparison of results, for a population of galaxies}\label{section: sample comparison}

\begin{figure}
    \centering
    \includegraphics[width=\linewidth]{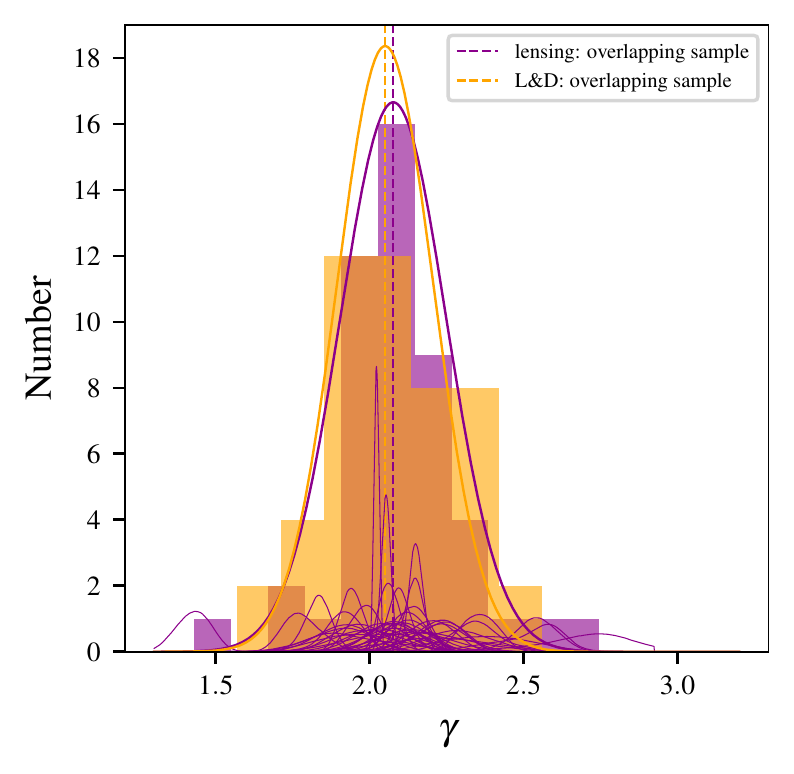}
    \caption{Logarithmic slopes $\gamma$ of galaxies' total mass density $\rho(r)\propto r^{-\gamma}$, measured using Lensing-only and Lensing\,\&\,Dynamics techniques, for a common ``overlapping'' sample of 48 galaxies. The two high Gaussian curves and dashed lines illustrate the best-fit mean and intrinsic scatter fitted via Equation~\ref{likelihood}. Lower curves show the posterior PDFs of individual lensing measurements, to illustrate their additional measurement uncertainty.}
    \label{Figure: distribution slope method comp}
\end{figure}

\begin{figure}
    \centering
    \includegraphics[width=\linewidth]{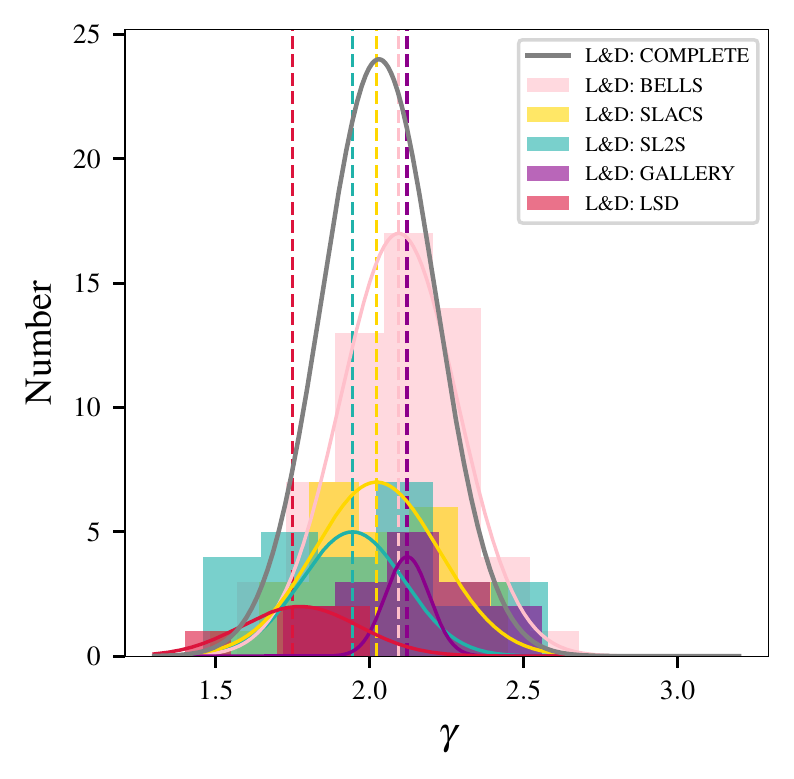}
    \caption{Logarithmic total mass-density slopes measured using the lensing\,\&\,dynamics method for the ``complete'' sample of 123 galaxies, but split into the SLACS, BELLS, GALLERY, SL2S, and LSD samples. Coloured curves show the best-fit mean and intrinsic scatter of galaxies in each survey, calculated as in figure~\ref{Figure: distribution slope method comp}. The grey curve shows the best fit to all 123 galaxies. }
    \label{Figure: distribution slopes dynamics}
\end{figure}

We assume that each individual galaxy's mass-density slope $\gamma_i$ belongs to an underlying Gaussian distribution of slopes with mean $\langle\gamma\rangle$ and intrinsic scatter $\sigma_\gamma$. The likelihood function of $\langle\gamma\rangle$ and $\sigma_\gamma$ is 
\begin{equation}\label{likelihood}
    \mathscr{L}(\langle\gamma\rangle, \sigma_\gamma|\{\gamma_i\}) =  \prod_{i}\frac{\mathrm{exp\Big[-\frac{(\gamma_i-\langle\gamma\rangle)^2}{2(\sigma_\gamma^2+\sigma^2_{\gamma_{i}})}\Big]}}{\sqrt{2\pi(\sigma_\gamma^2+\sigma^2_{\gamma_{i}})}},
\end{equation}
where $\sigma_{\gamma_i}$ is the uncertainty on the individual slope measurements $\gamma_i$. One can then infer the posterior probability distribution function (PDF) of $\langle\gamma\rangle$ and $\sigma_\gamma$ using Bayes' theorem
\begin{equation}\label{bayes}
    p(\langle\gamma\rangle, \sigma_\gamma|\{\gamma_i\}) \propto p(\langle\gamma\rangle, \sigma_\gamma) \mathscr{L}(\langle\gamma\rangle, \sigma_\gamma|\{\gamma_i\}), 
\end{equation}
where $p(\langle\gamma\rangle, \sigma_\gamma)$ is the prior. We assume uniform priors on $\langle\gamma\rangle$ and $\sigma_\gamma$ and fit for them using the nested sampling algorithm \texttt{dynesty} \citep{Speagle2019} via an implementation using the probabilistic programming language \texttt{PyAutoFit} \citep{pyautofit}. Note that lensing-only analysis uses a non-linear fitting procedure that yields asymmetric and non-Gaussian uncertainties on lensing slopes $\sigma_{\gamma_i}$. We approximate these as a split normal distribution, i.e.\ Gaussian uncertainty with
\begin{align}
    \sigma_{\gamma_i}&=\sigma_{\gamma_i}^\textrm{ue} \qquad  \textrm{if}  \qquad \gamma_i < \langle\gamma\rangle, \\
    \sigma_{\gamma_i}&=\sigma_{\gamma_i}^\textrm{le} \qquad   \textrm{if} \qquad \gamma_i >\langle\gamma\rangle,
\end{align}
where $\sigma_{\gamma_i}^\textrm{ue}$ and $\sigma_{\gamma_i}^\textrm{le}$ correspond to the upper and lower uncertainties at the 68.7$\%$ credible region of the individual $\gamma^\textrm{lensing}$ PDF. 

For galaxies in the overlapping sample, we measure mean logarithmic density slope $\langle\gamma\rangle=2.075_{-0.024}^{+0.023}$ and intrinsic scatter $\sigma_\gamma=0.172^{+0.022}_{-0.032}$ (purple curve in figure~\ref{Figure: distribution slope method comp}), errors quoted are at the 68\% credible region. This is consistent with \LDfull measurements from the literature for the same sample, $\langle\gamma^\textrm{\LD}\rangle=2.050_{-0.031}^{+0.023}$ and $\sigma_\gamma^\textrm{\LD}=0.156^{+0.030}_{-0.026}$ (orange curve in figure~\ref{Figure: distribution slope method comp}).

Robustly for different methods, we thus confirm a slightly super-isothermal distribution of mass around galaxies in our overlapping sample. This is consistent with \cite{Auger2010}'s original Lensing \& Dynamics analysis of the entire SLACS sample, $\langle\gamma\rangle^\textrm{\LD}= 2.078\pm0.027$ and $\sigma_\gamma^\textrm{\LD}=0.16\pm0.02$, which has been verified in repeat analyses \citep{Ruff2011, Li2018}. We confirm that this result is also reproduced in an analysis of \LD measurements for our complete sample $\langle\gamma^\textrm{\LD}\rangle= 2.030^{+0.019}_{-0.020}$ and $\sigma_\gamma^\textrm{\LD}=0.184^{+0.020}_{-0.019}$ (grey curve in figure~\ref{Figure: distribution slopes dynamics}).

\begin{figure}
    \centering
    \includegraphics[width=\linewidth]{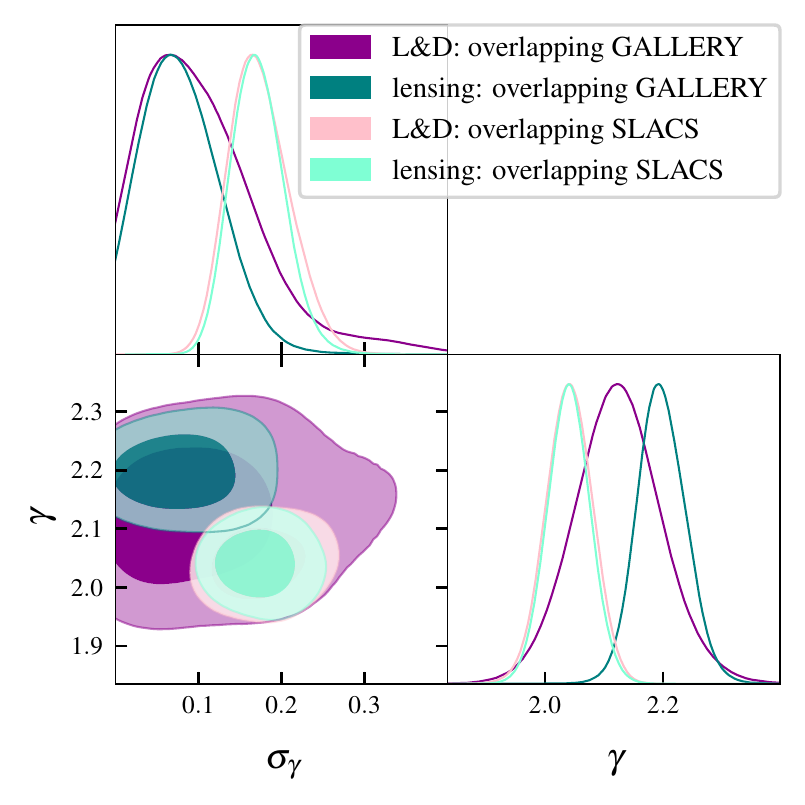}
    \caption{Posterior PDFs from fitting the mean $\gamma$ and intrinsic scatter $\sigma_\gamma$ of the SLACS and GALLERY overlapping sample of lenses assuming a Gaussian parent distribution. The GALLERY lenses have on average steeper density slopes than SLACS with both the lensing and dynamics (L\&D) and lensing only approach. }
    \label{Figure: mean and intrinsic scatter method comp}
\end{figure}

Splitting our complete sample into its parent surveys (figure~\ref{Figure: distribution slopes dynamics}), we note that the (high redshift) GALLERY lenses are the only sample with a mean logarithmic slope steeper than the (low redshift) SLACS sample. This remains true for the sub-samples of SLACS and GALLERY lenses that go into our overlapping sample. The posterior PDF contours in figure~\ref{Figure: mean and intrinsic scatter method comp} show that lenses in the GALLERY sample have steeper slopes with smaller intrinsic scatter, for both lensing-only (dark green) and \LDfull (dark purple) measurements, than for the SLACS sample (light green and purple). Something may be unusual in the selection technique used to find GALLERY lenses (see Sections~\ref{section: evolution} and ~\ref{section: benefits} for further discussion).

\subsection{Comparison of results, for individual galaxies}\label{section: individual comparison}

\begin{figure}
    \centering
    \includegraphics[width=\linewidth]{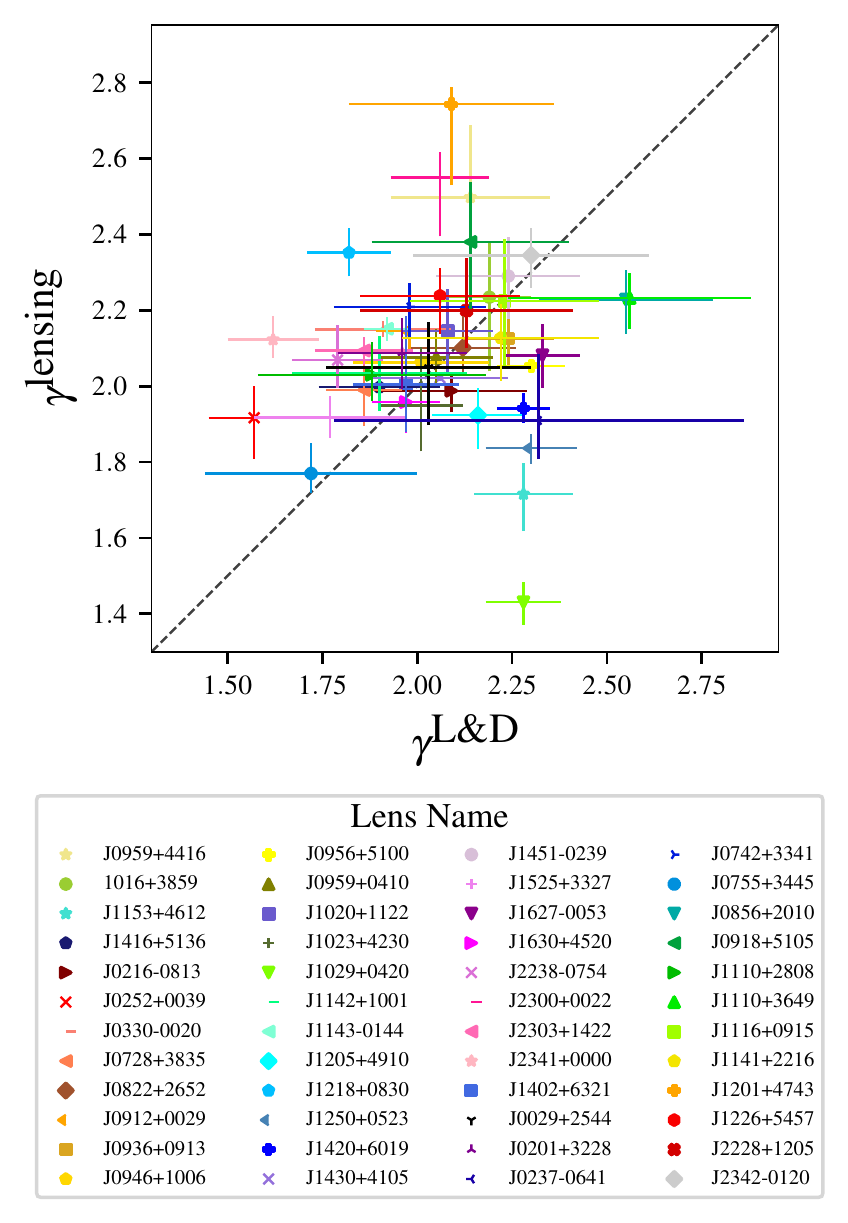}
    \caption{Logarithmic mass-density slopes of 48 individual galaxies in our ``overlapping'' sample, measured using lensing and stellar kinematics $\gamma^\mathrm{\LD}$ or lensing-only $\gamma^\textrm{lensing}$ methods. The identity line is plotted solely to guide visualisation. In the legend 33 SLACS lenses are listed first, followed by 15 GALLERY lenses. }
    \label{Figure: slope method comp}
\end{figure}

To further test whether the lensing-only and \LDfull methods are measuring the same total mass-density slopes, Figure~\ref{Figure: slope method comp} compares their measurements for each of the 48 galaxies in the overlapping sample. Assuming the measurement errors are correct, we investigate to what extent the true underlying slope measurements for this sample of galaxies are correlated. To do this we assume that the combination of $\gamma^\textrm{lensing}$ and $\gamma^\textrm{L\&D}$ can be described by a bi-variate Gaussian distribution with likelihood
\begin{equation}\label{likelihood bivariate}
    \mathscr{L}(\mathbf{\mu}, \mathbf{\Sigma}_\mathrm{int} | \mathbf{\gamma}_i) = \prod_{i}\frac{\mathrm{exp}\Big(-\frac{1}{2}(\mathbf{\gamma}_i-\mathbf{\mu})^T \mathbf{\Sigma}_i^{-1} (\mathbf{\gamma}_i-\mathbf{\mu})\Big)}{\sqrt{(2\pi)^2|\mathbf{\Sigma}_i|}},
\end{equation}
where $\mathbf{\mu} = [\langle\gamma^{\textrm{lensing}}\rangle, \langle\gamma^{\textrm{L\&D}}\rangle]$ is the vector mean, $\mathbf{\Sigma}_i = \mathbf{\Sigma}_\textrm{int}+ \mathbf{\Sigma}_{\textrm{err, i}}$ is the covariance matrix, and $\mathbf{\gamma}_i=[\gamma_i^{\textrm{lensing}},\gamma_i^{\textrm{L\&D}}]$ are the individual slope measurements. The vector mean $\mathbf{\mu}$ and the covariance matrix 
\begin{equation}
    \mathbf{\Sigma}_\mathrm{int} = 
    \begin{bmatrix}
    (\sigma^\mathrm{lensing}_{\gamma})^2 & \rho\sigma^\mathrm{lensing}_{\gamma}\sigma^\mathrm{\LD}_{\gamma} \\
    \rho\sigma^\mathrm{lensing}_{\gamma}\sigma^\mathrm{\LD}_{\gamma} &(\sigma^\mathrm{\LD}_{\gamma})^2
    \end{bmatrix},
\end{equation}
together describe the intrinsic distribution of the lensing-only and \LD slopes, where $\rho$ is the intrinsic correlation between $\gamma^\textrm{lensing}$ and $\gamma^\textrm{\LD}$. We assume the two measurement errors are uncorrelated such that the covariance matrix $\mathbf{\Sigma}_\textrm{err, i}$ is given by 
\begin{equation}
\mathbf{\Sigma}_\textrm{err, i}=
    \begin{bmatrix}
    \left( \sigma^\mathrm{lensing}_{\gamma,i} \right)^2 & 0 \\
    0 &\left(\sigma^\mathrm{\LD}_{\gamma,i} \right)^2
    \end{bmatrix},
 \end{equation}
 where $\sigma^\mathrm{lensing}_{\gamma,i}$ and $\sigma^\mathrm{\LD}_{\gamma,i}$ are the individual measurement errors on $\gamma^\textrm{lensing}$ and $\gamma^\textrm{\LD}$, respectively. Note that in this case we approximate the asymmetric lensing-only measurement errors as Gaussian with $\sigma_{\gamma,i}=(\sigma_{\gamma,i}^\textrm{ue}+\sigma_{\gamma,i}^\textrm{le})/2$.
 
Using Bayes' theorem (equation~\ref{bayes}) we infer the PDFs of the independent parameters $\langle\gamma^{\textrm{lensing}}\rangle$, $\langle\gamma^{\textrm{\LD}}\rangle$, $\sigma^\mathrm{lensing}_{\gamma}$, $\sigma^\mathrm{L\&D}_{\gamma}$, and $\rho$ in equation~\ref{likelihood bivariate}. We fit for these parameters with an MCMC sampling process using the Python implementation \texttt{emcee} \citep{Foreman_Mackey_2013}. The means $\langle\gamma^\textrm{lensing}\rangle=2.085^{+0.031}_{-0.030}$ and $\langle\gamma^\textrm{L\&D}\rangle=2.050^{+0.034}_{-0.033}$, and intrinsic scatters $\sigma_\gamma^\textrm{lensing}=0.191^{+0.027}_{-0.023}$ and $\sigma_\gamma^\textrm{L\&D}=0.159^{+0.031}_{-0.027}$ inferred with this bi-variate model agree with those fitted separately in Section~\ref{section: sample comparison}. We infer a correlation co-efficient $\rho=-0.150^{+0.223}_{-0.217}$, consistent with no correlation at the 68\% credible region. At 2$\sigma$ confidence the model implies a wide range of correlation coefficients, both negative and positive (-0.554 - 0.276), are consistent with the data. With this level of measurement uncertainty, we cannot definitively detect a correlation between the slopes measured with lensing and those measured with \LD. There is, however, no obvious systematic offset between the two methods: the mean difference is $\langle\gamma^\textrm{lensing}-\gamma^\textrm{\LD}\rangle=0.031\pm0.042$ and the data points appear scattered randomly either side of the identity line.

%% file: 4_Correlations.tex
\subsection{Correlations with the total-mass density slope}\label{section: correlations}
{\renewcommand{\arraystretch}{1.4}
\begin{table*}
\centering 
\begin{tabular}{c | r | r r | r r | r r } 
\hline\hline
\multirow{2}{*}{Covariate ($x$)} & \multirow{2}{*}{$\langle x\rangle$} &  \multicolumn{2}{c|}{Gradient ($\frac{\partial\langle\gamma\rangle}{\partial x})$} & \multicolumn{2}{c|}{Intercept ($\langle\gamma_0\rangle$)} & \multicolumn{2}{c}{Scatter ($\sigma_\gamma$)}\\
\cline{3-8}
& & $\gamma^\textrm{lensing}$ & $\gamma^\mathrm{\LD}$ & $\gamma^\textrm{lensing}$ & $\gamma^\mathrm{\LD}$ & $\gamma^\textrm{lensing}$ & $\gamma^\mathrm{\LD}$ \\
\hline 
$z_\textrm{lens}$ & 0.319    &   $0.248^\textrm{+0.174}_\textrm{-0.178}$ &   $0.043^\textrm{+0.215}_\textrm{-0.224}$ &  $2.077^\textrm{+0.029}_\textrm{-0.028}$ &  $2.058^\textrm{+0.031}_\textrm{-0.039}$ &  $0.173^\textrm{+0.026}_\textrm{-0.022}$ &  $0.165^\textrm{+0.034}_\textrm{-0.028}$ \\
$R_\textrm{eff}$ & 7.27     &   $0.017^\textrm{+0.009}_\textrm{-0.009}$ &    $-0.02^\textrm{+0.01}_\textrm{-0.013}$ &  $2.076^\textrm{+0.024}_\textrm{-0.027}$ &  $2.054^\textrm{+0.024}_\textrm{-0.029}$ &   $0.169^\textrm{+0.027}_\textrm{-0.02}$ &  $0.155^\textrm{+0.029}_\textrm{-0.033}$ \\
$\textrm{log}[M_\textrm{tot}]$ & 11.2 &  $0.254^\textrm{+0.082}_\textrm{-0.088}$ &   $-0.12^\textrm{+0.118}_\textrm{-0.108}$ &  $2.076^\textrm{+0.032}_\textrm{-0.029}$ &  $2.054^\textrm{+0.038}_\textrm{-0.037}$ &  $0.165^\textrm{+0.026}_\textrm{-0.022}$ &  $0.158^\textrm{+0.033}_\textrm{-0.028}$ \\
$^{R_\textrm{Ein}}/_{R_\textrm{eff}}$ & 0.91  & $-0.036^\textrm{+0.025}_\textrm{-0.023}$ &  $-0.026^\textrm{+0.046}_\textrm{-0.041}$ &  $2.076^\textrm{+0.027}_\textrm{-0.026}$ &  $2.049^\textrm{+0.036}_\textrm{-0.033}$ &  $0.173^\textrm{+0.031}_\textrm{-0.024}$ &  $0.162^\textrm{+0.024}_\textrm{-0.026}$ \\
$\sigma_\textrm{e2}$ & 260 &     $0.001^\textrm{+0.0}_\textrm{-0.0}$ &   $0.002^\textrm{+0.001}_\textrm{-0.001}$ &   $2.072^\textrm{+0.03}_\textrm{-0.024}$ &  $2.073^\textrm{+0.032}_\textrm{-0.025}$ &  $0.172^\textrm{+0.024}_\textrm{-0.023}$ &  $0.138^\textrm{+0.033}_\textrm{-0.021}$ \\
$\textrm{log}[\Sigma_\textrm{tot}]$ & 9.6   &  $-0.304^\textrm{+0.167}_\textrm{-0.152}$ &   $0.631^\textrm{+0.221}_\textrm{-0.285}$ &  $2.078^\textrm{+0.024}_\textrm{-0.028}$ &   $2.049^\textrm{+0.03}_\textrm{-0.025}$ &  $0.167^\textrm{+0.024}_\textrm{-0.019}$ &  $0.131^\textrm{+0.033}_\textrm{-0.027}$ \\

\hline 
\end{tabular}
\caption{Correlations between galaxies' total mass-density slopes $\gamma$ and other galaxy observables, as visualised in Figure~\ref{Figure: correlations}. Uncertainties are the $1\sigma$ credible regions on the gradient, intercept, and scatter on the covariate $x$, as in Equation~\eqref{Equation: x variable}.}\label{Table: correlations}
\end{table*}}

\begin{figure}
    \centering
    \includegraphics[width=\linewidth, trim={0 2mm 0 0},clip]{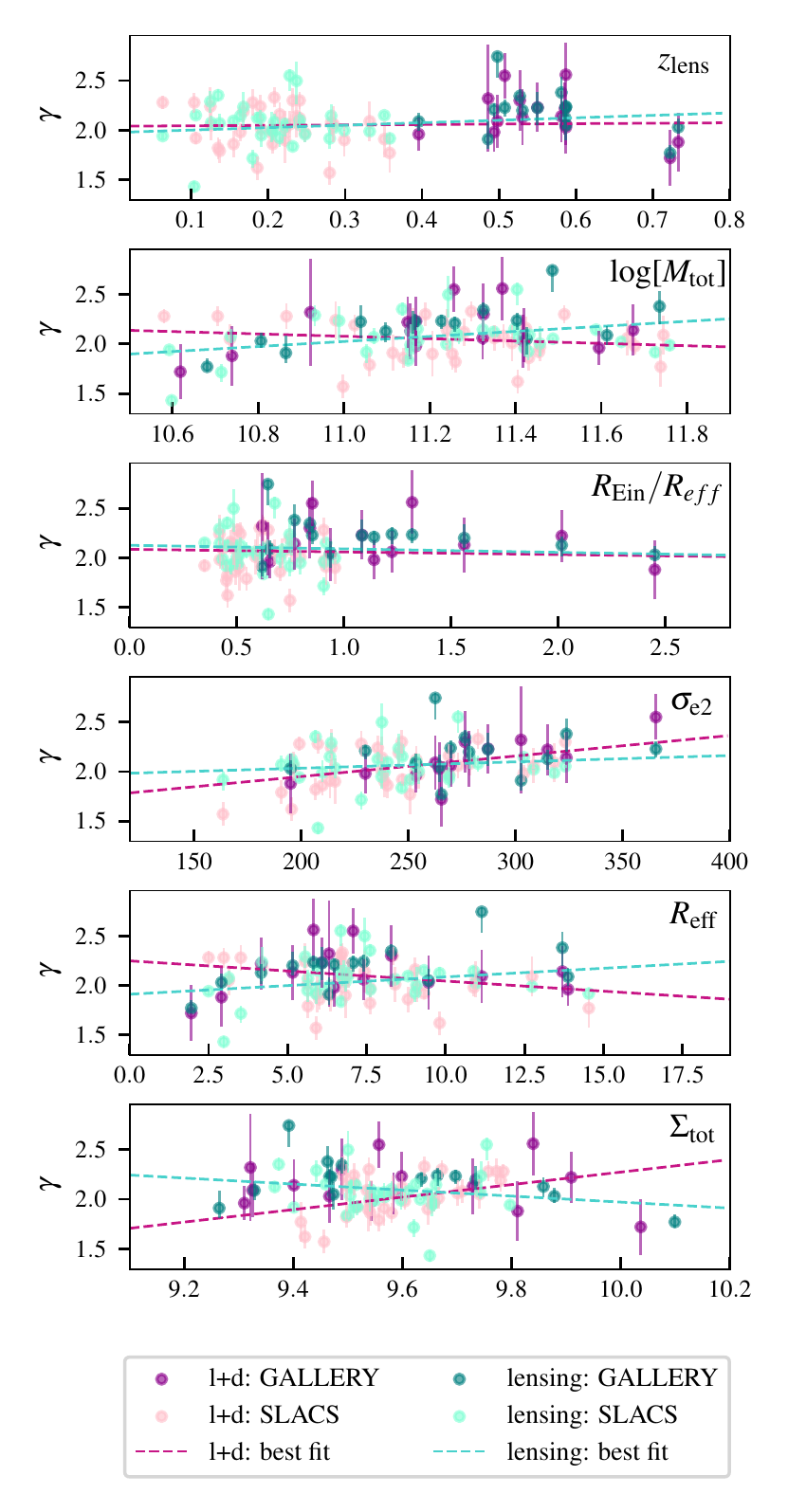}
    \caption{Correlation between total mass-density slopes $\gamma$ and other properties of a lens galaxy. Measurements with lensing-only or \LDfull techniques are consistent, except for the correlation with total mass density $\Sigma_\mathrm{tot}$. Numerical parameters of the best-fit lines are listed in Table~\ref{Table: correlations}.}
    \label{Figure: correlations}
\end{figure}

Since it is difficult to quantify for individual galaxies whether the (lensing-only and \LDfull) methods are measuring the same slope, we instead investigate whether they infer the same global dependence with other observable properties of galaxies. We continue to assume that the slopes are drawn from a parent Gaussian distribution, but we now assume the mean of the distribution (in Equation~\ref{likelihood}) is described by 
\begin{equation} \label{Equation: x variable}
    \langle\gamma\rangle(x) = \langle\gamma\rangle_0+\frac{\partial\langle\gamma\rangle}{\partial x}(x-\langle x\rangle),
\end{equation}
where $x$ is the galaxy observable. The free parameters in the model are now the mean slope $\langle\gamma\rangle_0$ at the average of the chosen galaxy observable $\langle x\rangle$, the intrinsic scatter of the distribution of slopes $\sigma_\gamma$, and the linear correlation coefficient $\frac{\partial\langle\gamma\rangle}{\partial x}$. We again use {\tt dynesty} to fit these free parameters, in successive analyses where $x$ represents the redshift of the lens galaxy, its effective radius, its total mass (equation~\ref{eq: M total}), its normalised Einstein radius, its velocity dispersion, or its total surface mass density. The best fit parameters from this procedure are listed in Table~\ref{Table: correlations} and visualised in Figure~\ref{Figure: correlations}\footnote{We additionally looked for correlations of $\gamma$ with lens light axis-ratio, mass model axis-ratio and external shear magnitude, but found no evidence for a correlation in either the lensing-only or lensing and dynamics measurements.}. 

For most galaxy observables, we find correlations with the logarithmic density slope that are consistent (at 2$\sigma$ confidence) for the lensing-only and \LD analyses. The only exception is the dependence upon total surface mass density. The \LD slopes imply a positive correlation of $\frac{\partial\langle\gamma\rangle}{\partial \Sigma_\textrm{tot}}=0.631^{+0.434}_{-0.492}$ (2$\sigma$ uncertainty), which is in agreement with previous \LD slope measurements for larger samples \citep{Auger2010, Sonnenfeld2013a}; whereas the lensing slopes imply zero or slightly negative correlation $\frac{\partial\langle\gamma\rangle}{\partial \Sigma_\textrm{tot}}=-0.304^{+0.358}_{-0.357}$. This may indicate that the methods are measuring different slopes, and we shall investigate this further in Section~\ref{sigma tot model}. 

We find that $\langle\gamma^\mathrm{\LD}\rangle$ has non-zero correlation (at $>2\sigma$ statistical significance) with only two lens observables: total mass density,  and velocity dispersion. Whereas, $\langle\gamma^\mathrm{lensing}\rangle$ has non-zero correlation with only total mass. Given that our analysis does not account for uncertainty on the $x$ variable, we caution that these coefficients may be overestimated -- particularly for the dependence with velocity dispersion, for which typical measurements have $\sim10\%$ uncertainty.

%% file: 4_Redshift.tex
\section{Dependence of the total-mass density slope on redshift}\label{multiple covariates}

The logarithmic density slope of mass in a galaxy is governed by the relative amounts of stellar and dark matter, and the physical processes that modify their distribution as the galaxy evolves. Studying how $\langle\gamma\rangle$ depends upon redshift can therefore constrain universal models of galaxy formation.\footnote{As emphasised by \cite{Sonnenfeld2013b}, these measurements represent how the population mean density slope depends on the population parameters of the galaxies included in the model, and not how the mass-density slope evolves for an individual galaxy. By combining their measurements with literature values for the evolution of the mass and size of early-type galaxies, \cite{Sonnenfeld2013b} measured the average redshift evolution of an individual galaxy to be consistent with zero $\frac{d\gamma^{\LD}}{dz} = 0.10\pm0.12$.}

To quantify the dependence of the mean density profile slope on redshift, it is necessary to account for any other confounding variables by including them as covariates in the model. We investigate variables that are well motivated from previous \LD analyses \citep{Sonnenfeld2012, Auger2010, Li2018}, the total surface mass density and normalised Einstein radius. These two variables are strongly correlated (with a Pearson correlation coefficient of 0.66), to the degree that including both of them as covariates would yield degenerate and unphysical coefficients. Therefore we fit only one of these covariates at a time, modelling the mean logarithmic density slope 
\begin{equation}\label{eq: two covariates}
    \langle\gamma\rangle(z_\textrm{lens},x)=\langle\gamma\rangle_0+\frac{\partial\langle\gamma\rangle}{\partial z}(z_\textrm{lens}-0.3)+\frac{\partial\langle\gamma\rangle}{\partial x}(x-\langle x\rangle),
\end{equation}
where the free parameters to be fitted are $\langle\gamma\rangle_0$, the mean slope at $z_\textrm{lens}=0.3$ and $x=\langle x\rangle$, as well as $\frac{\partial\langle\gamma\rangle}{\partial z}$ and $\frac{\partial\langle\gamma\rangle}{\partial x}$, the linear coefficients of covariates $z_\textrm{lens}$ and $x$. We again perform fits using the nested sampling algorithm \texttt{dynesty} via the probabilistic programming language \texttt{PyAutoFit}. We assume uniform priors on gradient parameters $\frac{\partial\langle\gamma\rangle}{\partial z}$ and $\frac{\partial\langle\gamma\rangle}{\partial x}$ between $-10$ and $10$, and uniform priors on the intercept $\langle\gamma\rangle_0$ between $0$ and $5$.

\subsection{Allowing for dependence on surface mass density}\label{sigma tot model}

Previous studies have shown that a galaxy's logarithmic density slope measured using \LD \citep{Auger2010a, Sonnenfeld2013a} correlates with both its total and stellar surface mass density. In Section~\ref{section: correlations} we confirmed this for the \LD slopes but found that lensing-only slopes were consistent with zero correlation at 2$\sigma$. We investigate whether this discrepancy persists when we fit the density slopes of galaxies in the overlapping sample, but allowing for simultaneous variation with both redshift and total mass density (equation~\eqref{eq: two covariates}). When fitting to lensing-only results, we use covariate $x=\Sigma_\textrm{tot}^\textrm{lensing}$, which uses $R_\textrm{Ein}^\textrm{PL}$ in equation~\eqref{eq: sigma tot}. When fitting to \LD results, we use covariate $x=\Sigma_\textrm{tot}^{\LD}$, the power law density profiles inferred by L\&D analyses in the literature, but using $R_\textrm{Ein}^\textrm{SIE}$ in equation~\eqref{eq: sigma tot}. 

\begin{figure}
    \centering
    \includegraphics[width=\linewidth]{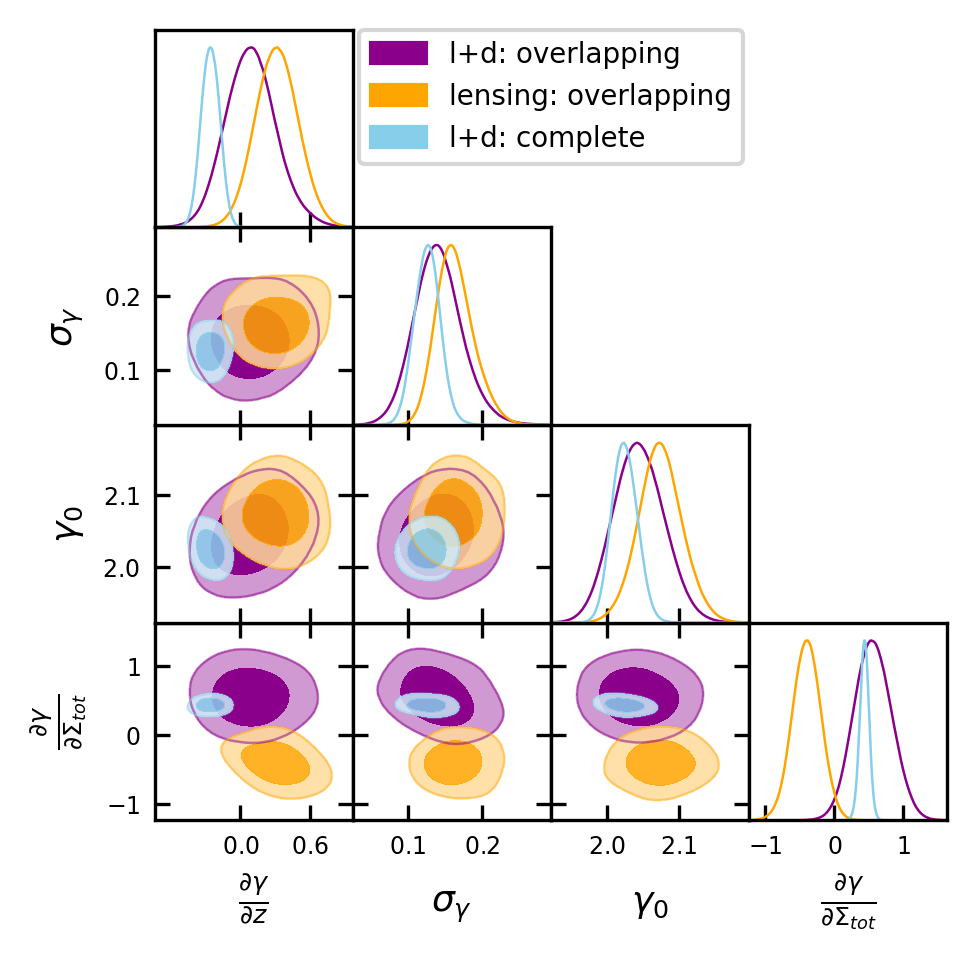}
    \caption{The 68\% (dark) and 95\% (light) marginalised confidence limits on posterior probabilities of the mean, intrinsic scatter, and linear coefficients for the dependence of slope on redshift and total surface mass density for both lensing-only  (orange contours) and lensing\&dynamics slopes (purple contours).}
    \label{Figure: mass density model}
\end{figure}

Best-fit parameters for lensing-only and L\&D analyses of the overlapping sample are listed in Table~\ref{Table: sigma tot model}, and the full posterior probability distributions are shown in Figure~\ref{Figure: mass density model}. The coefficient for variation with redshift is consistent between the two methods at 2$\sigma$. Surprisingly, however, our lensing-only analysis suggests that $\frac{\partial\langle\gamma\rangle}{\partial z} = 0.345^\text{+0.144}_\text{-0.167}$ is greater than zero at 2$\sigma$ confidence. \LD analysis of the same galaxies implies that $\frac{\partial\langle\gamma^{\LD}\rangle}{\partial z} = 0.045^\text{+0.217}_\text{-0.177}$ is consistent with zero. Fitting the complete sample of \LD slopes (blue contours in  Figure~\ref{Figure: mass density model}) yields a value less than zero at 4$\sigma$ confidence $\frac{\partial\langle\gamma^{\LD}\rangle}{\partial z} = -0.259^\text{+0.084}_\text{-0.082}$ \citep[in better agreement with  measurements in the literature;][see Table~\ref{Table: literature values}]{Auger2010, Bolton2012, Sonnenfeld2013a, Li2018}.

Coefficients describing the dependence of density slope on surface mass density are inconsistent between lensing-only and \LD analyses. For the overlapping sample of galaxies, the lensing-only coefficient is $\frac{\partial\langle\gamma\rangle}{\partial\Sigma_\textrm{tot}} = -0.432^\text{+0.175}_\text{-0.191}$, while \LD suggests $\frac{\partial\langle\gamma^{\LD}\rangle}{\partial\Sigma_\textrm{tot}} = 0.659^\text{+0.250}_\text{-0.264}$. Note that these results come from reasonably small populations of galaxies, and may be subject to outliers. For the complete lensing only sample, the increase in sample size leads to correlation coefficients with both redshift and surface mass density that are consistent with zero at 2$\sigma$ confidence (see Table~\ref{Table: sigma tot model}). Nonetheless, the coefficient with $\Sigma_\textrm{tot}$ remains inconsistent with those inferred for both the complete and overlapping \LD samples.

\subsection{Allowing for dependence on the radius where measurements are made}\label{r ratio model}

\begin{figure}
    \centering
    \includegraphics[width=\linewidth]{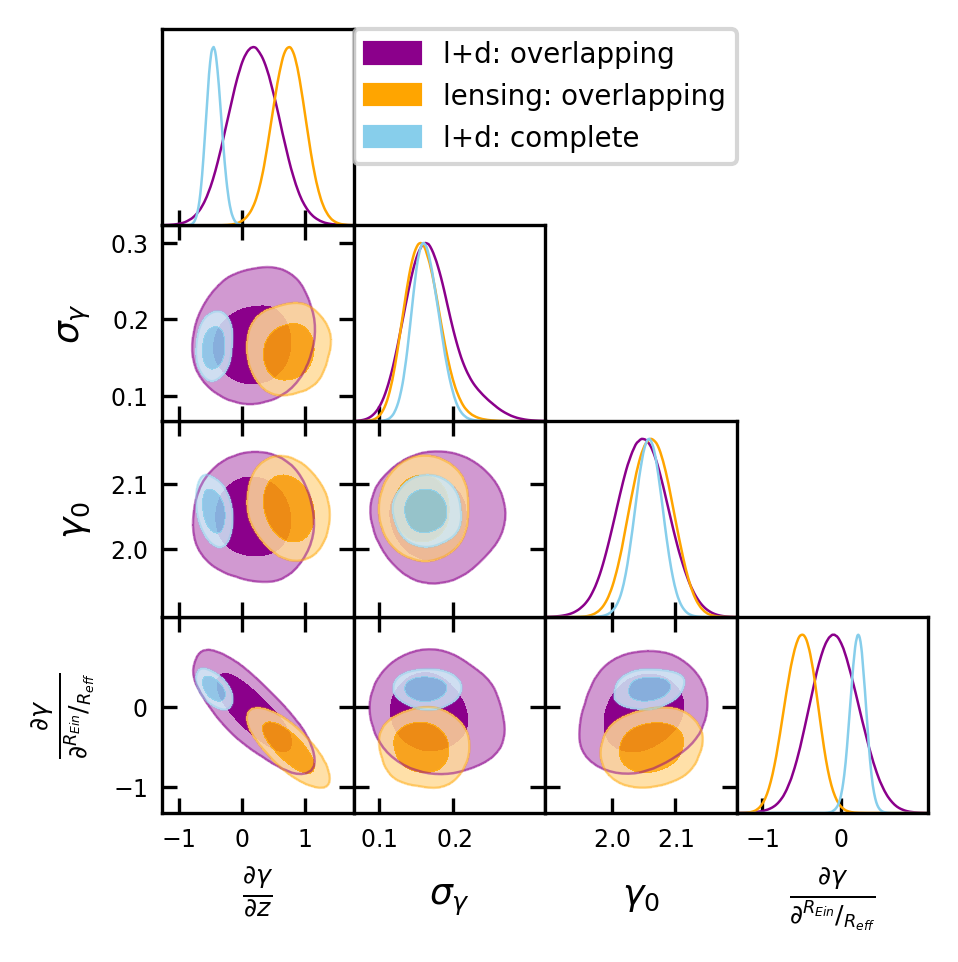}
    \caption{The 68\% (dark) and 95\% (light) marginalised confidence limits on posterior probabilities of the mean, intrinsic scatter, and linear coefficients for the dependence of slope on redshift and normalised Einstein radius for both lensing-only (orange contours) and lensing\&dynamics slopes (purple contours).}
    \label{Figure: r ratio model}
\end{figure}

If measurements of the density slope are sensitive to the radius at which the measurement is constrained, this could bias our inference about redshift dependence of the mean slope. Because the normalised Einstein radius  $^{\textrm{R}_{\textrm{Ein}}}/_{\textrm{R}_{\textrm{eff}}}$ typically increases with redshift for geometric reasons, one should simultaneously fit variation with $^{\textrm{R}_{\textrm{Ein}}}/_{\textrm{R}_{\textrm{eff}}}$ and $z_\textrm{lens}$ so as to not bias either result. Indeed, \cite{Li2018} demonstrated that \LD slopes display an increasing trend with radius, whilst still inferring a negative trend with redshift, for the BELLS, GALLERY, and SL2S samples. We now fit the two-covariate model (Equation~\ref{eq: two covariates}), with $x=\textrm{R}_{\textrm{Ein}}/\textrm{R}_{\textrm{eff}}$. Best-fit parameters for lensing-only and L\&D analyses of the overlapping sample are listed in Table~\ref{Table: r ratio model}, and the full posterior probability distributions are shown in Figure~\ref{Figure: r ratio model}.

Best-fit parameters of the lensing-only and L\&D models are consistent at 2$\sigma$ confidence for the overlapping sample. Albeit, for logarithmic density slopes measured with a lensing-only analysis, we infer relationships with redshift $\frac{\partial\langle\gamma\rangle}{\partial z} = 0.812^\textrm{+0.252}_\textrm{-0.263}$ and normalised Einstein radius $\frac{\partial\langle\gamma\rangle}{\partial \textrm{R}_{\textrm{Ein}}/\textrm{R}_{\textrm{eff}}} = -0.539^\textrm{+0.160}_\textrm{-0.191}$ at over 2$\sigma$ confidence, whereas the \LD inference $\frac{\partial\langle\gamma\rangle}{\partial z} = -0.002^\textrm{+0.303}_\textrm{-0.265}$ and $\frac{\partial\langle\gamma\rangle}{\partial \textrm{R}_{\textrm{Ein}}/\textrm{R}_{\textrm{eff}}} = 0.074^\textrm{+0.194}_\textrm{-0.294}$ is consistent with no correlation at 2$\sigma$. Note that, as for the $\Sigma_\textrm{tot}$ model, the coefficients for the complete lensing-only sample are both consistent with zero (see Table~\ref{Table: r ratio model}). The complete \LD sample infers coefficients $\frac{\partial\langle\gamma\rangle}{\partial z} = -0.428^\textrm{+0.119}_\textrm{-0.127}$ and $\frac{\partial\langle\gamma\rangle}{\partial \textrm{R}_{\textrm{Ein}}/\textrm{R}_{\textrm{eff}}} = 0.21^\textrm{+0.114}_\textrm{-0.100}$ that are consistent with measurements in the literature.  

{\renewcommand{\arraystretch}{1.5}
\begin{table*}
\centering 
\begin{tabular}{l | c | c | c | c} 
\hline\hline
Sample & $\langle\gamma\rangle_0$ & $\sigma_\gamma$ & $\frac{\partial\langle\gamma\rangle}{\partial z}$ & $\frac{\partial\langle\gamma\rangle}{\partial \Sigma_\textrm{tot}}$ \\
\hline
complete: LD & $2.024^\text{+0.019}_\text{-0.021}$ & $0.126^\text{+0.023}_\text{-0.015}$ & $-0.259^\text{+0.084}_\text{-0.082}$ &  $0.423^\text{+0.068}_\text{-0.075}$ \\
overlapping: LD & $2.051^\text{+0.031}_\text{-0.037}$ & $0.127^\text{+0.028}_\text{-0.024}$ &  $0.045^\text{+0.217}_\text{-0.177}$ &   $0.659^\text{+0.250}_\text{-0.264}$ \\
overlapping: lensing & $2.071^\text{+0.027}_\text{-0.026}$ & $0.159^\text{+0.028}_\text{-0.018}$ &  $0.345^\text{+0.144}_\text{-0.167}$ & $-0.432^\text{+0.175}_\text{-0.191}$ \\
complete: lensing & $2.097^\text{+0.032}_\text{-0.029}$ & $0.202^\text{+0.023}_\text{-0.023}$ &  $0.147^\text{+0.174}_\text{-0.154}$ & $-0.225^\text{+0.181}_\text{-0.151}$ \\

\hline 
\end{tabular}
\caption{Best-fit values of free parameters in a two-covariate model (Equation~\ref{eq: two covariates}) describing the correlation between galaxies' logarithmic density slope, $\gamma$, with redshift $z$ and total surface mass density $\Sigma_\mathrm{tot}$. The total surface mass density is calculated inside the effective radius of a de Vaucouleurs fit to the stellar emission (equation~\ref{eq: sigma tot}). All errors are quoted at $2\sigma$ confidence intervals.}\label{Table: sigma tot model}
\end{table*}}

{\renewcommand{\arraystretch}{1.5}
\begin{table*}
\centering 
\begin{tabular}{l | c | c | c | c} 
\hline\hline
Sample & $\langle\gamma\rangle_0$ & $\sigma_\gamma$ & $\frac{\partial\langle\gamma\rangle}{\partial z}$ & $\frac{\partial\langle\gamma\rangle}{\partial R_\textrm{Ein}/_{R_\textrm{eff}}}$ \\
\hline
complete: LD & $2.056^\textrm{+0.023}_\textrm{-0.022}$ & $0.163^\textrm{+0.016}_\textrm{-0.019}$ & $-0.428^\textrm{+0.119}_\textrm{-0.127}$ &     $0.210^\textrm{+0.114}_\textrm{-0.100}$ \\
overlapping: LD & $2.053^\textrm{+0.033}_\textrm{-0.028}$ & $0.165^\textrm{+0.027}_\textrm{-0.025}$ & $-0.002^\textrm{+0.303}_\textrm{-0.265}$ &  $0.074^\textrm{+0.194}_\textrm{-0.294}$ \\
overlapping: lensing & $2.063^\textrm{+0.028}_\textrm{-0.023}$ & $0.156^\textrm{+0.027}_\textrm{-0.021}$ &  $0.812^\textrm{+0.252}_\textrm{-0.263}$ &  $-0.539^\textrm{+0.160}_\textrm{-0.191}$ \\
complete: lensing &  $2.095^\textrm{+0.03}_\textrm{-0.029}$ & $0.199^\textrm{+0.027}_\textrm{-0.025}$ &  $0.331^\textrm{+0.216}_\textrm{-0.228}$ & $-0.276^\textrm{+0.176}_\textrm{-0.156}$ \\

\hline 
\end{tabular}
\caption{Best-fit values of free parameters in a two-covariate model (Equation~\ref{eq: two covariates}) describing the correlation between galaxies' logarithmic density slope, $\gamma$, with redshift $z$ and normalised Einstein radius $^{R_\mathrm{Ein}}/_{R_\mathrm{eff}}$. The effective radius values $R_\mathrm{Eff}$ are literature values} of de~Vaucouleurs fits to the stellar emission. All errors are quoted at $2\sigma$ confidence intervals.\label{Table: r ratio model}
\end{table*}}

%% file: 5_Discussion.tex
\section{Discussion and Comparison With Previous Studies}\label{Discussion}

\subsection{Bulge-halo conspiracy?}
That the total mass-density profiles of massive elliptical galaxies is nearly isothermal has been observed in X-ray emission \citep{Humphrey2006}, dynamical modelling \citep{Serra2016, Poci2017}, and lensing and dynamical analyses \citep{Koopmans2006, Barnabe2009, Auger2010, Sonnenfeld2013a}. Given that neither the stellar nor dark matter components are individually described by a single power law, this remarkable observation about their sum is known as the ``bulge-halo consipracy''. On average, taking into account current measurement uncertainties, our analysis is consistent with this result. By fitting to only the imaging data of a sample of 48  strong lenses from the SLACS and GALLERY surveys, we measure slightly super-isothermal total-mass density slopes, with mean $\langle\gamma\rangle=2.075^{+0.023}_{-0.024}$ and intrinsic scatter $\sigma_\gamma=0.172^{+0.022}_{-0.032}$. Previous \LD analyses of exactly the same galaxies yield consistent measurements $\langle\gamma^\textrm{\LD}\rangle=2.050^{+0.023}_{-0.031}$ and  $\sigma^\textrm{\LD}_\gamma=0.156^{+0.030}_{-0.026}$.

If the true density profiles of massive elliptical lens galaxies are indeed power-law distributions, then one would expect a perfect correlation between the slopes constrained with the different methods. For a sample of 21 SLACS systems analysed using a similar lensing-only method, \cite{Shajib2021} were unable to detect a correlation between slopes measured using lensing only and \LD. They measured a bi-weight mid-correlation of $0.01\pm0.16$, where the errors on the correlation coefficient were calculated from the 68\% confidence interval of the coefficients calculated from 1000 random draws of their lensing-only and \LD slopes from the posterior PDFs. With more than double the number of systems, if we adopt the same approach as \cite{Shajib2021}, we continue to find no correlation between the lensing and L\&D slopes (bi-weight mid-correlation $0.08^{+0.11}_{-0.12}$). Moreover, using our own approach that takes into account the covariance between the intrinsic distributions of slopes (described in Section~\ref{section: individual comparison}), we measure a correlation coefficient of $-0.150^{+0.223}_{+0.217}$, suggesting an even wider range of correlation coefficients are consistent with the data. Therefore, although we can not rule out the existence of a global power law given the measurement uncertainties, the lack of an obvious correlation between the slopes measured using the different methods may be indicating that some of the systems deviate from a strict power law. 

\subsection{Are the lensing and dynamics and lensing-only methods constraining the same quantity?}\label{section: toy model}

\begin{figure}
\centering
\includegraphics[width=0.9\columnwidth]{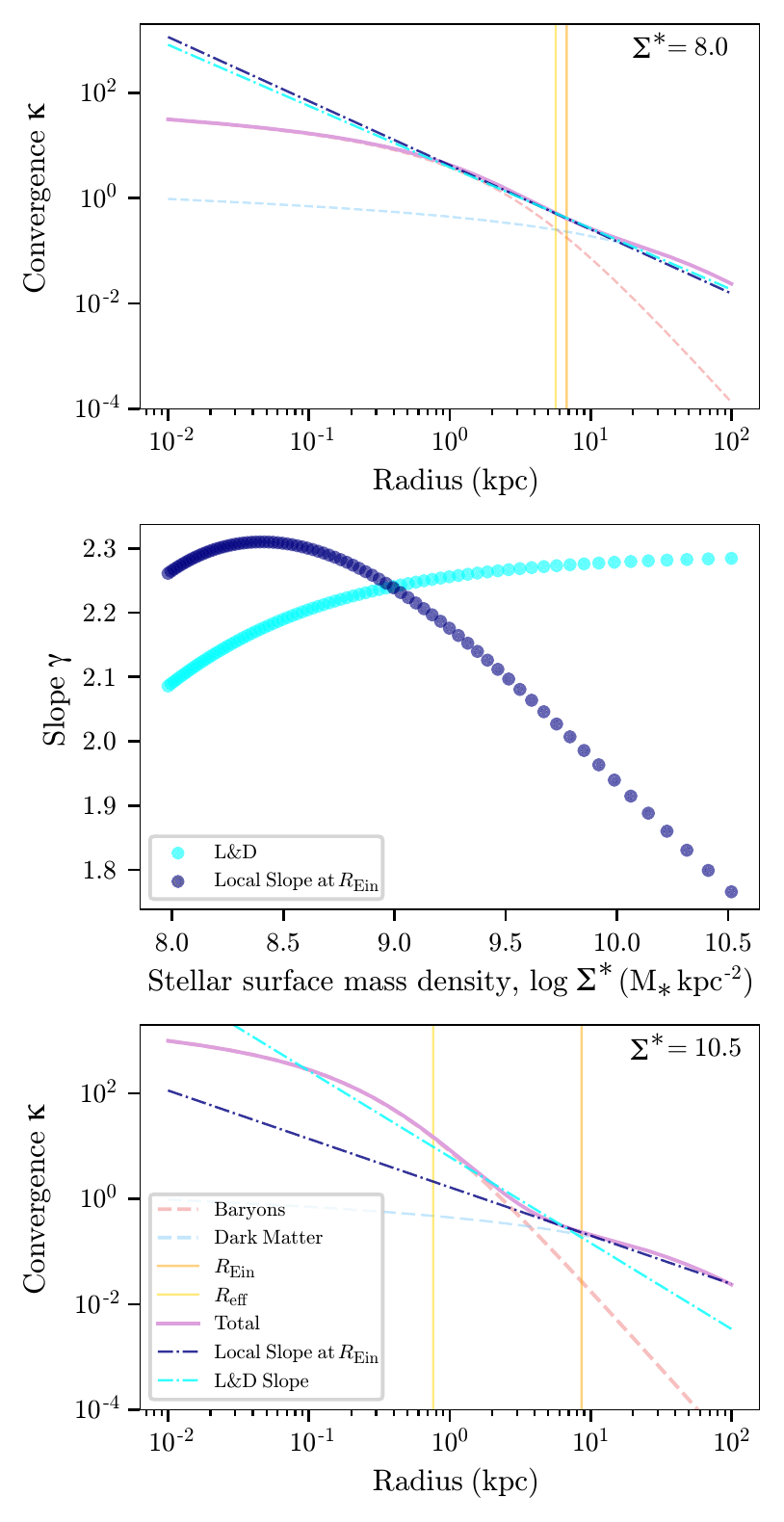}
\caption{Analytic model of an azimuthally-symmetric lens, which reproduces and explains behaviour observed in the data. The distribution of mass is described as the sum of Hernquist (stellar) and NFW (dark matter) profiles. As the stellar surface mass density increases from the top to bottom panel, the slope of the total mass-density profile constrained between the Einstein radius and effective radius (cyan dash-dotted line) steepens, mimicking the lensing and dynamics positive relationship with surface mass density. Conversely, the slope constrained locally at the Einstein radius (dark blue dash-dotted line) flattens, reproducing the negative relationship with surface mass density observed for the slopes measured using lensing only. The top and bottom panel represent the mass density-profiles, and their fitted \LD and local slopes, for the first and last points plotted in the middle panel which shows how these quantities behave as the stellar surface mass density increases.}
\label{toy model}
\end{figure}

Although lensing-only and \LD analyses yield consistent {\it mean} values of logarithmic density slopes for a population of galaxies, this does not necessarily imply that the two analyses constrain the same quantity for each {\it individual} galaxy. Lensing-only analyses are sensitive to the profile at the Einstein radius \citep[e.g.][]{Koopmans2006, Treu2010, Suyu2017}, whereas \LD analyses (combining measurements of velocity dispersion and Einstein radius) probe the integrated profile between the effective and Einstein radii.
If galaxies' total density profiles deviate from a power law, measurements of the logarithmic slope at different radii will yield different results. \cite{Shankar2017} report a connection between the observed dependence of $\gamma^{\LD}$ on stellar mass and effective radius (hence stellar surface mass density) and the relative amounts of stellar and dark matter in the region of the mass density profile that is being probed. They find that steeper $\gamma^{\LD}$ are inferred from the inner profile where the stellar component steepens. Similarly, with models of early-type galaxies built from analytic stellar and dark matter profiles, \cite{DuttonTreu2014} showed that the strength of the correlation between $\gamma^{\LD}$ and dark matter fraction largely determines the strength of the correlation between $\gamma^{\LD}$ and stellar density (among other galaxy variables).

We found in Section~\ref{multiple covariates} that measurements of a galaxy's logarithmic density slope can be made independently of most of its observable properties. The main complications are caused by variations in a galaxy's total surface mass density, $\Sigma_\textrm{tot}$. For a lensing-only analysis, we obtain negative values of $\frac{\partial\langle\gamma\rangle}{\partial \Sigma_\textrm{tot}}$, while \LD analyses are consistently positive (Table~\ref{Table: sigma tot model}). A negative coefficient seemingly runs counter to the expectation (also demonstrated with stellar kinematics methods \citealt{Poci2017}) that galaxies with higher stellar densities have higher central densities, and hence steeper total mass-density slopes. This disagreement may therefore indicate that the \LD and lensing-only methods are constraining different quantities. Indeed, it is notable that the multivariate model we fit to the lensing-only analyses (overlapping and complete) with normalised Einstein radius as a second covariate, infer similar coefficients $\frac{\partial\langle\gamma\rangle}{\partial\textrm{R}_{\textrm{Ein}}/\textrm{R}_{\textrm{eff}}}$ to the coefficients $\frac{\partial\langle\gamma\rangle}{\partial \Sigma_\mathrm{tot}}$ in the models fitted with surface mass density as a covariate (see Tables~\ref{Table: r ratio model} and ~\ref{Table: sigma tot model}). Since these quantities are strongly correlated, and can therefore not be fitted for simultaneously, it is difficult to interpret the coefficients individually. It may be that the negative relationship we infer with total surface mass density is a consequence of a more fundamental dependence on the radius at which the lensing slope is measured.

We now put forward a toy model that can explain the observed behaviour. We first construct a model of the distribution of mass in an early type galaxy, comprising baryons in a spherically-symmetric Hernquist profile, and dark matter in a spherically-symmetric NFW profile (Figure~\ref{toy model}). In line with previous studies \citep{DuttonTreu2014, Shankar2017}, we found Salpeter-like values were necessary to produce values of total-mass density slope that were representative of the \LD observations. We fix the total stellar mass (at 11.64 log[$M^*/M_\odot$]), then adjust the NFW parameters to obtain a dark matter fraction within half the effective radius representative of observations of early-type galaxies (these are small such that at this radius the stellar and total masses, hence surface densities, are similar). Following \cite{DuttonTreu2014} we then approximate the \LD mass-density slope measurement as the mass-weighted density slope within the effective radius 
\begin{equation}
   \gamma^{\textrm{L\&D}}_{\textrm{proxy}}(r)= \frac{1}{M(<\mathrm{R_\mathrm{eff}})}\int^{\mathrm{R_\mathrm{eff}}}_0 -\gamma(r)4\pi r^2\rho(r) dr,
\end{equation}
where $M(<\mathrm{R_\mathrm{eff}})$ is the total mass within the effective radius, and $-\gamma(r)\equiv d\textrm{log}\rho/d\textrm{log}r$ is the local logarithmic slope of the given density profile $\rho(r)$. We then assume that the lensing-only method measures the local logarithmic slope at the Einstein radius. In this model, increasing the stellar surface mass-density from $\Sigma^*=8$\ log[$M_\odot \textrm{kpc}^{-2}$] (left panel), to $\Sigma^*=10.5$\ log[$M_\odot \textrm{kpc}^{-2}$] (right panel), by decreasing the effective stellar radius, raises the inferred \LD slope (cyan dot-dashed line) from 2.08 to 2.28 but reduces the inferred local logarithmic slope at the Einstein radius (navy dot dashed line) slope from 2.26 to 1.76 -- similar to our observations of real galaxies.

The negative relationship of the local slope at $\textrm{R}_\textrm{Ein}$ in this model (middle panel Figure~\ref{toy model}) appears to occur at larger values of normalised Einstein radius. The effective radius is typically in a heavily baryon-dominated regime; the Einstein radius near an inflection point in the total density profile, created by the transition from baryon- to dark matter-domination. As we increase the stellar density, the steepening stellar profile strengthens the inflection point (deviating further from a power-law), and the Einstein radius moves out farther towards the inflection point and a shallower slope. However, all this is averaged over by a \LD measurement. That we observe the same behaviour in real galaxies suggests their total density profile might also contain a detectable inflection point. If further work supports this hypothesis, that the inflection is detected by a lensing-only measurement, but averaged over by a \LD measurement, future analyses that combine measurements may be able to constrain deviations from the bulge-halo conspiracy.

If lensing-only measurements are near an inflection point, we anticipate this would produce an increased intrinsic scatter compared to the L\&D measurements. This is because the local nature of the lensing measurement depends more sensitively on the inflection point, whereas the L\&D measurement averages over the extended inner radial density. Table~\ref{Table: sigma tot model} shows that for the overlapping lens sample, when redshift $z$ and total surface mass density $\Sigma_\mathrm{tot}$ are covariates, the lensing-only scatter is is $\sigma_\gamma=0.159^\text{+0.028}_\text{-0.018}$ and L\&D is $\sigma_\gamma=0.127^\text{+0.028}_\text{-0.024}$. As expected, the lensing-only measurement is higher, but they are consistent within $2\sigma$ confidence intervals. For the complete samples the lensing-only value is higher than the l\&D at over $2\sigma$ confidence, with values $\sigma_\gamma=0.202^\text{+0.023}_\text{-0.023}$ and $\sigma_\gamma=0.126^\text{+0.023}_\text{-0.015}$ respectively. This is tentative evidence the lensing-only method has more scatter, consistent with an inflection point, but a larger sample of overlapping lenses is necessary to confirm this.

We acknowledge that it is a strong assumption that the lensing only measurement constrains the local slope at the Einstein radius. This neglects the constraining power of the positions of the arcs in the image (i.e. the Einstein radius) that are fit for simultaneously with the gradients of the deflection angle field that constrain the slope in a real lensing analysis. If a lens's true underlying mass distribution is not a power-law, the inferred lensing-only slope measurement will be biased by the mass-sheet degeneracy (MSD) \citep{Schneider2013, Sluse2012}. The size of the bias depends on the difference in curvature of the true mass profile, near $R_{\rm Ein}$, compared to the fitted power-law \citep{Schneider2013a}. However, as discussed below, results from \citet{Cao2021} indicate that the MSD makes lensing-only measurement more closely trace the $L\&D$ measurement. 

Based on the MSD, \cite{Kochanek2020} emphasise that the only two quantities determined by lens data are the Einstein radius and the dimensionless and mass-sheet invariant quantity $\xi_2=\textrm{R}_\textrm{Ein}\alpha''(\textrm{R}_\textrm{Ein})/(1-\kappa_\textrm{Ein})$ where $\alpha''(\textrm{R}_\textrm{Ein})$ is the second derivative of the deflection profile at $\textrm{R}_\textrm{Ein}$. They argue that power law models have a one to one mapping between this quantity and the mass-density slope $\gamma=\xi_2/2+2$. With $\gamma$ calculated in this way for the Hernquist+NFW profiles plotted in Figure~\ref{toy model}, we do not find a negative relationship between total mass-density slope and stellar surface mass density. This may be implying that the mass-density profiles that make up our toy-model are too simplistic, that the way we induce an increase in stellar surface mass density is different to how this increase occurs in real galaxies, or that $\gamma=\xi_2/2+2$ does not well represent what we measure with lensing only in real galaxies. 

Understanding what slope lensing constrains when the underlying profile deviates from a power law will be invaluable in interpreting the results presented in this work. \cite{Cao2021} showed that the true profile's mass weighted slope within the Einstein radius better matched the total mass-density slope of lensing only fits to mock images, simulated with complex multiple Gaussian expansion + NFW profiles, than the mass weighted slope between 0.8 - 1.2 $\textrm{R}_\textrm{Ein}$. For these mock systems the mismatch between the power-law and the true density profiles can be compensated by a mass-sheet transformation (see Figure 8 of \citet{Cao2021}), which results in a fitted lensing only slope that resembles more closely the true density profiles' average slope over a local measurement as suggested in this work. Nonetheless, the 2$\sigma$ disagreement between the lensing only and \LD surface mass density coefficients implies a deviation of the underlying profile from a power law distribution, and the negative relationship of the lensing only slopes with normalised Einstein radius may well be the result of an inflection zone like that described in the toy-model put forward in this work. 

\subsection{Evolution of massive elliptical galaxies}\label{section: evolution}
{\renewcommand{\arraystretch}{1.5}
\begin{table*}
\centering 
\begin{tabular}{c | c | c | c | c | c | c} 
\hline\hline
Study & Samples & $\langle\gamma\rangle_0$ & $\sigma_\gamma$ & $\frac{\partial\langle\gamma\rangle}{\partial z}$ & $\frac{\partial\langle\gamma\rangle}{\partial \Sigma}$ & $\frac{\partial\langle\gamma\rangle}{\partial R_\textrm{Ein}/_{R_\textrm{eff}}}$\\
\hline
\cite{Bolton2012} & SLACS $\&$ BELLS & $2.11\pm0.02$ & $0.14\pm0.06$ & $-0.60\pm0.15$ & - & - \\
\cite{Sonnenfeld2013a} & SLACS, SL2S, $\&$ LSD & $2.08^{+0.02}_{-0.02}$ & $0.12^{+0.02}_{-0.02}$ & $-0.31^{+0.09}_{-0.10}$ & $0.38^{+0.07}_{-0.07}$ & - \\
\cite{Li2018} & SL2S, BELLS, $\&$ GALLERY & $1.981^{+0.024}_{-0.024}$ & $0.168^{+0.021}_{-0.017}$ & $-0.309^{+0.092}_{-0.083}$ & - & $0.194^{+0.092}_{-0.083}$ \\
\hline 
\end{tabular}
\caption{Comparison of the coefficients inferred for models in previous studies that have constrained the redshift dependence of lensing and dynamics total mass-density slopes. }\label{Table: literature values}
\end{table*}}

Although measurements of $\frac{\partial\langle\gamma\rangle}{\partial z}$ reflect the evolution of galaxy populations rather than individual galaxies, they can still inform models of the overall processes. For example, \cite{Shankar2018} found their observations could be reproduced only if the S\'ersic index of stellar components vary with redshift. Our \LD analysis confirms previous measurements in the literature, that galaxies' logarithmic density slopes decrease with redshift, i.e.\ they steepen with cosmic time (see Table~\ref{Table: literature values}). 
Interestingly, most cosmological simulations instead show a mild increase in density slopes with redshift \citep{wang2020, Wang2019, Remus2017}, which is inconsistent with \LD measurements but matches our lensing-only results. At present, it is not clear whether this discrepancy indicates a limitation of the simulations, systematics in the observations, or additional complexity in the physics, such as deviations from a power law profile. Notably, adjusting the method used to calculate density slopes in the Illustris simulation so it better represents observational techniques suggests a mild shallowing of slopes with redshift, $\frac{\partial\langle\gamma\rangle}{\partial z} = -0.03\pm0.01$ \citep{Xu2017}. Nonetheless, those authors caution that the method still suffers from systematic biases and does not account for sampling bias that will be present in the observational data. 

Galaxy selection effects are important. Both lensing-only and \LD analyses of our overlapping sample yield positive values of $\frac{\partial\langle\gamma\rangle}{\partial z}$ that do not match the results of larger samples (see Tables~\ref{Table: sigma tot model} and \ref{Table: r ratio model}). The positive coefficients are driven by the GALLERY lenses, which constitute most high redshift lenses in the overlapping sample, and have the steepest mean slopes. The unusual properties of GALLERY systems may even explain the differences between the lensing-only and \LD coefficients. Because the constraining power of \LD analyses degrade at high redshift (see Section~\ref{section: benefits}), the GALLERY sample does not contribute as much to the overall fit, and $\frac{\partial\langle\gamma\rangle}{\partial z}$ is not as significantly positive.

If lensing-only and \LD techniques measure different aspects of galaxies' mass distributions, as we suggested they might in Section~\ref{section: toy model}, it is unclear whether we should expect the dependence of these measurements on redshift to agree. Nevertheless, with the current level of statistical precision, the lensing-only and L\&D measurements are consistent when we model the same samples of lenses.

\subsection{Benefits of lensing-only analyses}\label{section: benefits}

Measurements using our lensing-only method do not degrade at high redshift. This is illustrated in Figure~\ref{Figure: err v redshift}, which compares the statistical uncertainty of lensing-only or \LD measurements of each slope, $\delta \gamma$, as a function of redshift. For the \LD analysis of  galaxies in the overlapping sample, we find a strong linear relationship between uncertainties and lens redshift, with slope 0.37$\pm$0.05 (1$\sigma$ errors, i.e.\ significant at $>$3$\sigma$). However, for lensing-only measurements, we measure much less degradation, with slope 0.06$\pm$0.04 (1$\sigma$ errors, i.e.\ consistent with zero at 2$\sigma$). Note that lensing-only measurements of (high redshift) GALLERY lenses have greater uncertainty than lensing-only measurements of (low redshift) SLACS lenses. This appears to be unrelated to the lens redshift, as we detect no correlation between measurement uncertainties and redshift for the SLACS and GALLERY samples separately. It is instead expected because the GALLERY lenses were selected due to Lyman-alpha emission from their source galaxies, which makes them inherently compact and less well resolved. 

Despite this selection effect, which disfavours only the lensing method, the lensing-only measurements of the GALLERY sample are better constrained than \LD measurements, which degrade due to increasing uncertainty on velocity dispersion measurements at high redshift. This highlights the potential of the lensing-only method to extend this analysis to higher redshifts without losing constraining power. Deviations from a global power law, as discussed in Section~\ref{section: toy model}, may complicate the interpretation of analyses like that presented in this study. In future, constraining how the parameters of more complex stellar plus dark matter distributions depend on redshift may be more appropriate to further our understanding of the evolution of ETGs, a problem well suited to strong lensing \citep{Nightingale2019}. \citet{Sonnenfeld2021} demonstrated the ability of strong lensing alone to calibrate stellar masses and constrain the inner dark matter density profile of galaxies with a hierarchical approach.

In the next couple of decades, lensing-only analyses could be possible at redshifts up to $z\sim2$, through surveys such as Euclid and the Vera Rubin Observatory that will discover large populations of high redshift lenses \citep{Collett2015}. Furthermore, the lensing-only measurements were constrained from the imaging data alone and can therefore scale to the hundreds of thousands of lenses that these surveys will observe, without requiring deep spectroscopic observations. Photometric redshifts \cite{Sonnenfeld2022} and fully-automated analyses \citep{Shajib2021, Etherington2022} will be key in this endeavour.

%% file: 6_Summary.tex
\section{Summary}\label{Summary}
We measure the distribution of mass around 48 early type galaxies in the SLACS and GALLERY strong lens surveys to test the `bulge-halo conspiracy' that stellar and dark matter together produce a power-law radial density profile with index $\gamma$. We compare two methods: a traditional Lensing \& Dynamics (\LD) technique that combines the Einstein radius from lensing with stellar kinematical data; and a lensing-only technique that fits every pixel in imaging data. The two methods yield consistent measurements of the parent distribution of $\gamma$. Our lensing-only technique finds a population average,  $\langle\gamma\rangle=2.075_{-0.024}^{+0.023}$,  with intrinsic scatter between galaxies of $\sigma_\gamma=0.172^{+0.022}_{-0.032}$

Two results hint at the fact that the conspiracy breaks down.
First, although the two methods yield consistent population-averaged measurements, they appear to differ for individual galaxies. If every galaxy has a single, well-defined power-law slope, it is surprising that we infer a statistically insignificant correlation coefficient of $-0.150^{+0.223}_{-0.217}$ although we cannot rule out a global power law with the current level of measurement uncertainty. Second, although both methods can measure $\gamma$ independently of most galaxy properties, measurements are correlated with total surface mass density (even when we fit multivariate models including redshift as a covariate). The lensing-only method yields a negative correlation, ${\partial\langle\gamma\rangle}/{\partial \Sigma_\textrm{tot}} = - 0.432^\text{+0.404}_\text{-0.348}$, whereas the \LD method yields a positive correlation, ${\partial\langle\gamma^{\LD}\rangle}/{\partial \Sigma_\textrm{tot}} = 0.659^\text{+0.481}_\text{-0.474}$.

We discuss a hypothesis that could explain these results. The \LD method measures the galaxy's mean density profile between the Einstein radius and its effective radius. This averages out deviations from a power law, pointing to a `bulge-halo conspiracy'. However, the lensing-only method is sensitive to the local slope at the Einstein radius. For galaxies in which the Einstein radius is larger than the effective radius, the Einstein radius typically occurs near the transition between the stellar-dominated core and the dark matter-dominated outskirts -- an inflection point where the total mass profile deviates from a power law. The inflection gets stronger as the stellar mass density increases.
Further studies \citep[e.g.][]{Cao2021, Kochanek2020} will be useful to test this hypothesis and to understand how deviations from a power law could affect previous inferences about galaxy evolution.

Any study of galaxy evolution must deal with selection effects. Our results suggest that galaxy redshift and stellar surface density affect the mass profile inferred with lensing methods (partly because they usefully change the Einstein radius, and so probe deviations from a power-law mass distribution). If selection effects can be understood, the lensing-only method will be able to analyse the large samples of lenses expected from surveys such as Euclid. This is because it requires only imaging data and can be automated. Its application will also be possible to higher redshifts than \LD techniques, whose statistical precision degrades with redshift due to uncertainties in obtaining accurate spectroscopy.

In this study we measured a redshift dependence of $\partial\langle\gamma\rangle/\partial z=0.345^{+0.322}_{-0.296}$ at fixed surface mass density for the lensing-only slopes, consistent with the same sample of \LD slopes but in tension with the complete \LD sample. A large sample of lenses from a single lens survey like Euclid will provide a tighter constraint on this conclusion and remove any biases from the different selection effects of the combined surveys. This will offer new insights into the formation and evolution galaxies out to redshift 2.0 and beyond.

\section*{Data Availability}
Tables containing the relevant data used for the analysis of each of the three observational samples of galaxies are available at \url{https://github.com/amyetherington/beyond_bulge_halo_data}.

%% file: software.tex
\section*{Software Citations}

This work uses the following software packages:

\begin{itemize}

\item
\href{https://github.com/astropy/astropy}{{Astropy}}
\citep{astropy1, astropy2}

\item
\href{https://github.com/dfm/corner.py}{{Corner.py}}
\citep{corner}

\item
\href{https://github.com/joshspeagle/dynesty}{{Dynesty}}
\citep{dynesty}

\item
\href{https://github.com/matplotlib/matplotlib}{{Matplotlib}}
\citep{matplotlib}

\item
\href{numba` https://github.com/numba/numba}{{Numba}}
\citep{numba}

\item
\href{https://github.com/numpy/numpy}{{NumPy}}
\citep{numpy}

\item
\href{https://github.com/rhayes777/PyAutoFit}{{PyAutoFit}}
\citep{pyautofit}

\item
\href{https://github.com/Jammy2211/PyAutoGalaxy}{{PyAutoGalaxy}}
\citep{pyautogalaxy}

\item
\href{https://github.com/Jammy2211/PyAutoLens}{{PyAutoLens}}
\citep{Nightingale2015, Nightingale2018, Nightingale2021}

\item
\href{https://www.python.org/}{{Python}}
\citep{python}

\item
\href{https://github.com/scikit-image/scikit-image}{{Scikit-image}}
\citep{scikit-image}

\item
\href{https://github.com/scikit-learn/scikit-learn}{{Scikit-learn}}
\citep{scikit-learn}

\item
\href{https://github.com/scipy/scipy}{{Scipy}}
\citep{scipy}

\item
\href{https://www.sqlite.org/index.html}{{SQLite}}
\citep{sqlite}

\end{itemize}